\documentstyle[psfig]{l-aa}

\def\spose#1{\hbox to 0pt{#1\hss}}
\def\simlt{\mathrel{\spose{\lower 3pt\hbox{$\mathchar"218$}}
     \raise 2.0pt\hbox{$\mathchar"13C$}}}
\def\simgt{\mathrel{\spose{\lower 3pt\hbox{$\mathchar"218$}}
     \raise 2.0pt\hbox{$\mathchar"13E$}}}

\begin{document}

\thesaurus{ }
\title{The Galactic Lithium Evolution Revisited}
\author{Donatella Romano$^{1,2}$, Francesca Matteucci$^{3,1}$, 
	Paolo Molaro$^2$, Piercarlo Bonifacio$^2$}
\offprints{D. Romano, e-mail: romano@sissa.it}
\institute{
SISSA/ISAS, 
Via Beirut, 2-4, 34014 Trieste, Italy 
\and
Osservatorio Astronomico di Trieste, 
Via G.B. Tiepolo, 11, 34131 Trieste, Italy 
\and
Dipartimento di Astronomia, Universit\`a di Trieste, 
Via G.B. Tiepolo, 11, 34131 Trieste, Italy}
\maketitle
\markboth{D. Romano et al.\,: The Galactic Lithium Evolution Revisited}{}

\begin{abstract}
The evolution of the $^7$Li abundance in the Galaxy has been computed by means 
of the two-infall model of Galactic chemical evolution. We took into account 
several stellar $^7$Li sources: novae, massive AGB stars, C-stars and Type II 
SNe. In particular, we adopted new theoretical yields for novae. We also took 
into account the $^7$Li production from GCRs. In particular, the absolute 
yields of $^7$Li, as suggested by a recent reevaluation of the contribution of 
GCR spallation to the $^7$Li abundance, have been adopted.

We compared our theoretical predictions for the evolution of $^7$Li abundance 
in the solar neighborhood with a new compilation of data, where we identified 
the population membership of the stars on a kinematical basis. A critical 
analysis of extant observations revealed a possible extension of the Li 
plateau towards higher metallicities (up to [Fe/H] $\sim$ $-$\,0.5 or even 
$-$\,0.3) with a steep rise afterwards.

We conclude that 1) the $^7$Li contribution from novae is required in order to 
reproduce the shape of the growth of A(Li) versus [Fe/H], 2) the contribution 
from Type II SNe should be lowered by at least a factor of two, and 3) the 
$^7$Li production from GCRs is probably more important than previously 
estimated, in particular at high metallicities: by taking into account GCR 
nucleosynthesis we noticeably improved the predictions on the $^7$Li abundance 
in the presolar nebula and at the present time as inferred from measures in 
meteorites and T\,Tauri stars, respectively. We also predicted a lower limit 
for the present time $^7$Li abundance expected in the bulge, a prediction 
which might be tested by future observations.
\keywords{Galaxy: chemical evolution -- stars: evolution, abundances -- 
         cosmic rays -- ISM: abundances plot}
\end{abstract}

\section{Introduction}

Since Spite \& Spite (1982) discovered that the oldest, warm halo dwarfs in 
the Galaxy all show almost the same $^7$Li abundance, several papers have 
appeared in the literature supporting their initial interpretation that this 
is the primordial abundance of $^7$Li (Spite, Maillard \& Spite, 1984; Spite 
\& Spite, 1986; Rebolo, Molaro \& Beckman, 1988, hereafter RMB; but see also 
Thorburn, 1994).\\
The substantial flatness of the plateau and the absence of intrinsic scatter 
(Spite et al.\,, 1996; Bonifacio \& Molaro, 1997), coupled with the detection 
of the fragile isotope $^6$Li in the metal-poor stars HD\,84937 (Smith et 
al.\,, 1993; Hobbs \& Thorburn, 1997; Cayrel et al.\,, 1999) and perhaps 
BD+42\,2667 (Cayrel et al.\,, 1999), are the arguments for claiming that no 
significant depletion mechanisms should have acted in these stars to modify 
the pristine $^7$Li abundance (but see also Ryan, Norris \& Beers, 1999).

We also note a competing theory, which claims that the highest $^7$Li 
abundance - measured in the most Li-rich Population I objects - is the 
primordial one (Boesgaard et al.\,, 1998). In this case some depletion 
mechanisms able to deplete Li in all the halo stars by the same amount 
acting on a Galactic lifetime timescale are required. Possible mechanisms 
include diffusion (Vauclair, 1988), rotational mixing (Pinsonneault et 
al.\,, 1992) and stellar winds (Vauclair and Charbonnel, 1995).

Recently, Fields \& Olive (1998), using a standard model of Galactic cosmic 
ray (GCR) nucleosynthesis, found that only little $^6$Li (and $^7$Li) 
depletion is allowed in halo stars. So that the observed Spite plateau should 
be indicative of the primordial Li abundance.

Younger stars span a wide range of lithium abundances. The highest values, 
measured in Orion T\,Tauri stars reach A(Li)\footnote{A(Li) = log$_{10}$
(N$_{^7Li}$/N$_{H}$) + 12.} $\sim$ 3.83 dex ($\sim$ 3.5 dex when corrected 
for NLTE effects, Duncan \& Rebull, 1996). Therefore, some mechanisms of 
$^7$Li production are required to increase the Li abundance from the plateau 
value to the present one. Nuclear reactions in stellar interiors and 
spallation processes on interstellar medium (ISM) particles involving either 
high or low energy GCRs have both been proposed as possible mechanisms able to 
synthesize $^7$Li and restore it back to the ISM, where it enters into the 
chemical composition of the new-formed stars.

Lithium evolution  has already been studied in detail by several authors. 
D'Antona \& Matteucci (1991), by means of a complete model of chemical 
evolution, have shown that both the Solar System lithium abundance and the 
rise from the Spite plateau could be explained assuming Li-production in 
classical novae and AGB stars. Later, novae were ruled out as lithium 
producers at a Galactic level (Boffin et al.\,, 1993). Matteucci et 
al.\,(1995) suggested a combination of $\nu$-process nucleosynthesis from 
Type II SNe and hot bottom burning in intermediate mass AGB stars to match 
the observations. Recently, Matteucci et al.\,(1999) used the same 
nucleosynthesis prescriptions to calculate the expected $^7$Li content in the 
Galactic bulge. Lithium production by low mass AGB stars (C-stars) and 
standard GCR nucleosynthesis were the key ingredients of the model of Abia et 
al.\,(1995), which was able to match also the behaviour of the lithium 
isotopic ratio. In any case, they had to require a percentage of Li-rich 
C-stars much higher (6\,--\,8\%) than observed (2\,--\,3\%). However, 
we note that C-stars at low metallicities are found much more numerous ($\sim$ 
10\%, Beers et al.\,, 1992; Norris et al.\,, 1997).

In this paper we deal with Galactic lithium evolution taking into account the 
lithium production both in stars and from GCRs. The aim of the paper is to 
reproduce the observed upper envelope of the diagram A(Li) vs [Fe/H], assuming 
that the Population II $^7$Li is the primordial one.\\
In \S 2 we present the data-set we have used to constrain the model results, 
in \S 3 we review the main candidates as stellar lithium producers, in \S 4 
we present the basic assumptions of the chemical evolution model and the 
lithium synthesis prescriptions, in \S 5 we illustrate our main results and 
in \S 6 we draw some conclusions.

\section{Observational data}

The observed evolution of the $^7$Li abundance with metallicity, when 
abundance determinations in disk and halo dwarfs are restricted to stars 
with effective temperature T$_{\mathrm{eff}}$ $\ge$ 5700 K, suggests that 
there is a general trend towards a larger ISM $^7$Li content with increasing 
metallicity. In previous papers (e.g. Matteucci et al.\,, 1995, 1999) the 
abundances used to constrain Galactic chemical evolution models were pointing 
to a smooth increase from the Spite plateau to the Solar System value. In this 
paper we show the results of an analysis performed on a large selection of 
data taken from the literature. Li detections for those stars which are 
tracing the upper envelope of the observational A(Li) vs [Fe/H] diagram have 
been critically analysed and a rise off the plateau steeper than previously 
assumed has been found. This result, coupled with a Spite plateau which 
extends towards larger metallicities, suggests a revision of the various 
contributions to the $^7$Li production from different sources. In particular, 
$^7$Li producers restoring their lithium to the ISM on long timescales should 
be preferred.  

$^7$Li measurements in stellar atmospheres have been selected from the 
literature in the metallicity range $-$\,1.5 $\le$ [Fe/H] $\le$ $-$\,0.5 dex
\footnote{Sources in the literature: Deliyannis et al.\,, 1990; Lambert et 
al.\,, 1991; Pilachowski et al.\,, 1993; Pasquini et al.\,, 1994; Spite et 
al.\,, 1996.}. This region is particularly interesting because it is the 
metallicity range where the observed lithium abundance is expected to start 
growing due to $^7$Li injection from the first Li factories.

From the point of view of stellar nucleosynthesis, it is important to 
determine the metallicity at which the Li abundance starts rising off the 
Spite plateau. For instance, the constancy of the Spite plateau at [Fe/H] 
$\le$ $-$\,1.5 translates into the requirement that Type II SNe coming from 
the first stellar generations should contribute only an amount of lithium
smaller than the primordial one.

Selection criteria have been applied in order to remove from the sample all 
those stars which are likely to have passed through phases during which they 
either destroyed or diluted their initial lithium content. To this purpose, 
we selected only stars not suffering any $^7$Li convective depletion, i.e. 
T$_{\mathrm{eff}}$ $\ge$ 5700 K, according to standard models. Stars with 
T$_{\mathrm{eff}}$ $<$ 5700 K are cool enough to be threatened by Li burning 
in deep convective envelopes already during the pre-main sequence (see also 
Ryan \& Deliyannis, 1998). Moreover, we retained only stars included in the 
HIPPARCOS catalogue which provides accurate information on kinematics and 
luminosities. Objects recognized as giants or subgiants are likely to show Li 
surface abundances affected by dilution and have been rejected.

The program stars are listed in Tab.3. For each star we give the HD, DM, G 
and HIP numbers and the U, V, W heliocentric space-velocity components with 
the associated errors. The membership to a specific Galactic population 
(halo or disk, either thin or thick) as derived from the kinematics and the 
evolutionary status are also provided.

\subsection{Evolutionary status}

In order to determine the evolutionary status of each star in our sample, we 
used the theoretical isochrones by Bertelli et al.\,(1994). We divided the 
program stars in three metallicity bins ([Fe/H] $\ge$ $-$\,0.75; $-$\,1.25 
$\le$ [Fe/H] $\le$ $-$\,0.75; and  [Fe/H] $\le$ $-$\,1.25). For each bin the 
data were compared with theoretical isochrones appropriately chosen for 
different ages at different metallicities. In the bin [Fe/H] $\ge$ $-$\,0.75 
we used the isochrones of 2\,--\,9 Gyr; in the bin $-$\,1.25 $\le$ [Fe/H] 
$\le$ $-$\,0.75 the 6\,--\,19 Gyr isochrones, and in the bin [Fe/H] $\le$ 
$-$\,1.25 those of 8\,--\,20 Gyr. The absolute magnitudes for the stars in 
our data-base were obtained from the HIPPARCOS parallaxes. The effective 
temperatures were calculated using the (B\,$-$\,V)\,--\,T$_{\mathrm{eff}}$ 
calibration of Alonso et al.\,(1996). Within each metallicity bin and by 
using the appropriate isochrone we distinguished among turn-off, giant and 
subgiant stars following the criteria outlined by Beers et al.\,(1990). We 
found that most of the stars are turn-off stars; only few among them 
(HIP\,36430, HIP\,37723, HIP\,86694, HIP\,103987, HIP\,115167, HIP\,116082) 
are subgiants and are, therefore, not considered in the analysis.

\subsection{Kinematics}

The heliocentric Galactic space-velocity components, U, V and W, calculated 
from the star's proper motion, parallax and radial velocity following the 
Johnson \& Soderblom (1987) analysis, are listed in columns 5, 6, 7 of Tab.3. 
The uncertainties $\sigma_{U}$, $\sigma_{V}$ and $\sigma_{W}$ are also 
given in columns 8, 9, 10. A left-handed coordinate system for U, V, W, so 
that they are positive in the directions of the Galactic anticenter, the 
Galactic rotation and the North Galactic Pole, respectively, is adopted. 
The radial velocities used to complement HIPPARCOS data are from the SIMBAD 
data-base or from the literature. For 12 objects lacking of radial velocity 
U, V, W are not provided. Stars for which the relative error on the parallax 
is greater or equal to 100\% ($\sigma_{\pi}/\pi \ge 1$) have also no 
kinematics determination.

With these data at hand we distinguished between a \emph{disk population} and 
a \emph{non-disk population}. Adopting a selection criterium based on the 
studies by Sandage \& Fouts (1987) and Beers \& Sommer-Larsen (1995), in 
order to belong to the disk population we required a star to have V $>$ 
$-$\,115 Km s$\,^{-1}$ and (U$^2$\,+\,W$^2$)$^{\frac{1}{2}}$ $<$ 150 Km 
s$\,^{-1}$ .\\
In Fig.1 we show the V vs (U$^2$\,+\,W$^2$)$^{\frac{1}{2}}$ diagram obtained 
for our data sample. Stars with metallicities below or above [Fe/H] = $-$\,1.0 
are shown with different symbols. In this plot the kinetic energy associated 
to the rotation around the Galactic center is compared to the kinetic energy 
associated to any other motion. As expected, stars with [Fe/H] $>$ $-$\,1.0 
concentrate in the region V $>$ $-$\,50 Km s$\,^{-1}$, 
(U$^2$\,+\,W$^2$)$^{\frac{1}{2}}$ $<$ 100 Km s$\,^{-1}$ and dilute outside. 
These objects  compose the bulk of the Galactic \emph{disk}. 
The \emph{thick-disk} should be envisaged in those stars which rotate more 
slowly and with larger random motions. For the sake of simplicity, we choose 
here to distinguish only between disk and non-disk stars.

Contrary to common assumption that all halo stars are also metal-poor we found 
three stars G\,170-156 ([Fe/H] = $-$\,0.8, HIP\,86321), G\,17-21 ([Fe/H] = 
$-$\,0.6, HIP\,80837) and G\,182-19 ([Fe/H] = $-$\,0.7, HIP\,86431) with 
metallicities larger than [Fe/H] $>$ $-$\,1.0 which do not show disk-like 
motion. One more star BD+01\,3421 ([Fe/H] = $-$\,0.5, HIP\,84905) is possibly 
belonging to the thick-disk.

On the other hand we found three stars HD\,166913 ([Fe/H] = $-$\,1.8, 
HIP\,89554), HD\,205650 ([Fe/H] = $-$\,1.3, HIP\,106749) and HD\,134169 
([Fe/H] = $-$\,1.6, HIP\,74079) with metallicity below $-$\,1.0 which possess 
a disk-like motion. These three stars are members of the metal-weak tail 
of the Galactic disk. Other two stars with these properties have been found 
by Bonifacio et al.\,(1999).

Five stars (HIP\,104659, HIP\,37789, HIP\,11952, HIP\,37853 and HIP\,55022) 
are lying just on the boundary which separates the halo from the disk stars.

In order to ascertain a possible correlation between the kinematics of the 
stars and the $^7$Li abundance, we drew in Fig.2 the graph A(Li)\,--\,[Fe/H] 
indicating with different symbols disk stars, non-disk stars and objects 
without a precise kinematical membership. From Fig.2 we note that the star 
HD\,160693 (HIP\,86431), with halo-like motion and a metallicity of $-$\,0.7, 
has A(Li) $<$ 1.2 falling  well below the Li plateau. On the other hand the 
metal-poor stars with disk motions show a Li abundance at the plateau level. 
Thus lithium depletion is likely related only to metallicity and independent 
from the kinematics.\\
These findings allow us to assume that the upper envelope of the observational 
diagram traces those stars which have suffered only a minor lithium depletion 
during their life. This interpretation is also supported  by the analysis of 
the stars with Be data, whose Be abundances imply Li abundances larger than 
actually observed (Molaro et al.\,, 1997). Therefore, the upper envelope 
results from the various lithium enrichment processes occurred during the 
overall Galactic lifetime.\\
Our revised compilation of the data from the literature shows a plateau which 
may extend at metallicities larger than previously assumed and points to a 
very steep rise off the plateau starting at [Fe/H] $\sim$ $-$\,0.5, $-$\,0.3. 
Details of the star by star critical Li analysis are provided in appendix.
\onecolumn
\begin{figure}[th!]
\psfig{figure=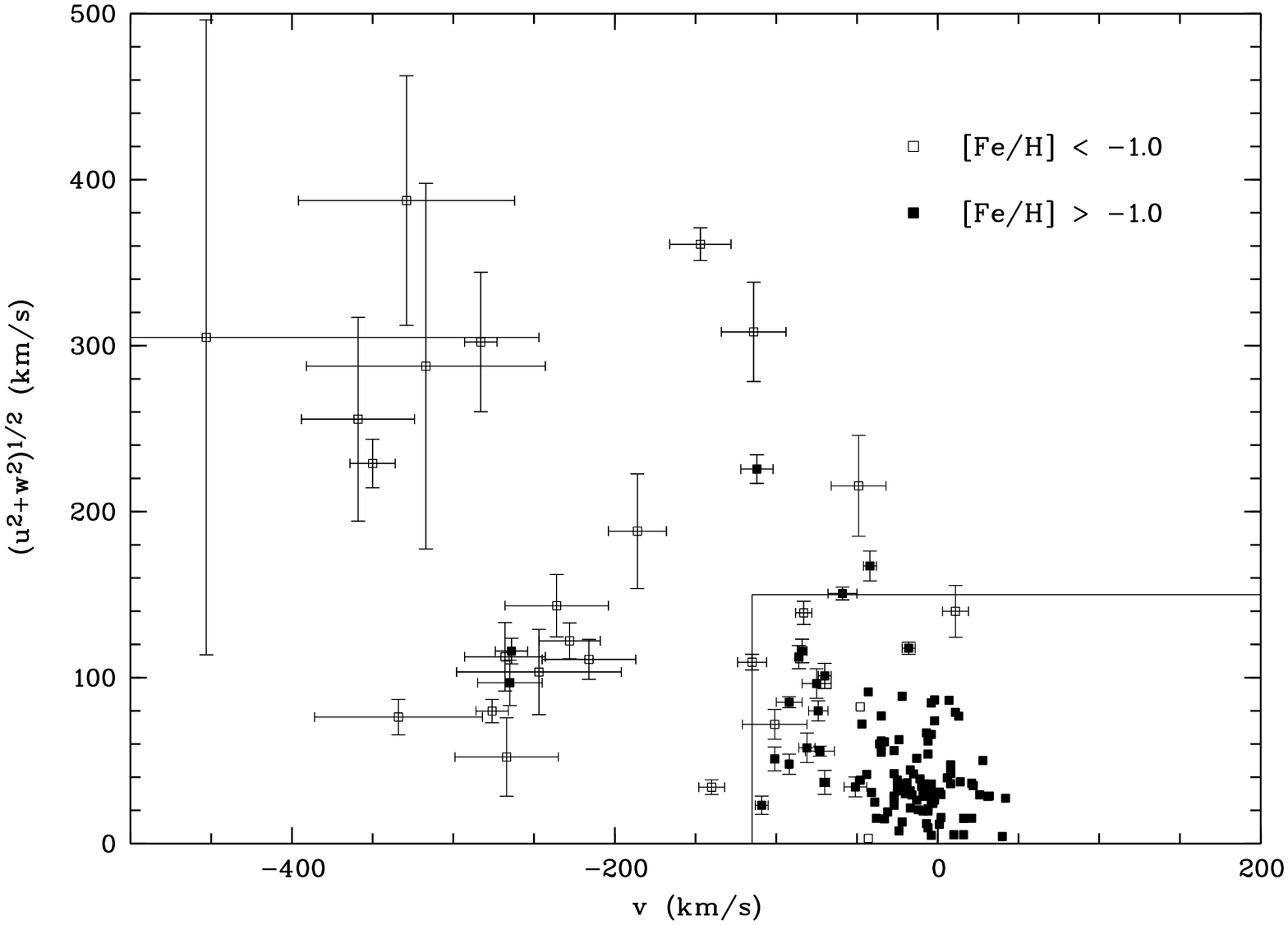,width=15cm,angle=-90}
\caption[]{\label{fig:fig 1} V versus (U$^2$\,+\,W$^2$)$^{\frac{1}{2}}$ 
diagram for the sample stars. The stars have been divided into two metallicity 
bins, on the grounds of the metallicity values given in Tab.4. In the case of 
multiple metallicity determinations we have adopted the average value. 
V $>$ $-$\,115 Km s$\,^{-1}$ and (U$^2$\,+\,W$^2$)$^{\frac{1}{2}}$ $<$ 150 Km 
s$\,^{-1}$ are the two selection criteria which have to be satisfied in order 
to ascribe the star to the Galactic disk (either thin or thick). For the sake 
of clarity, no error bars are drawn in the most dense region of the plot. In 
any case, stars lying in that region have quite small errors associated to the 
determinations of their velocity components. The stars HIP\,106749 (the empty 
square in the upper left corner of the box) and HIP\,74079 (the empty square 
in the lower left corner of the box) have been offset by +\,20 Km s$\,^{-1}$ 
in (U$^2$\,+\,W$^2$)$^{\frac{1}{2}}$ and by $-$\,40 Km s$\,^{-1}$ in V, 
respectively, to make them clearly visible.}
\end{figure}
\newpage
\begin{figure}[th!]
\psfig{figure=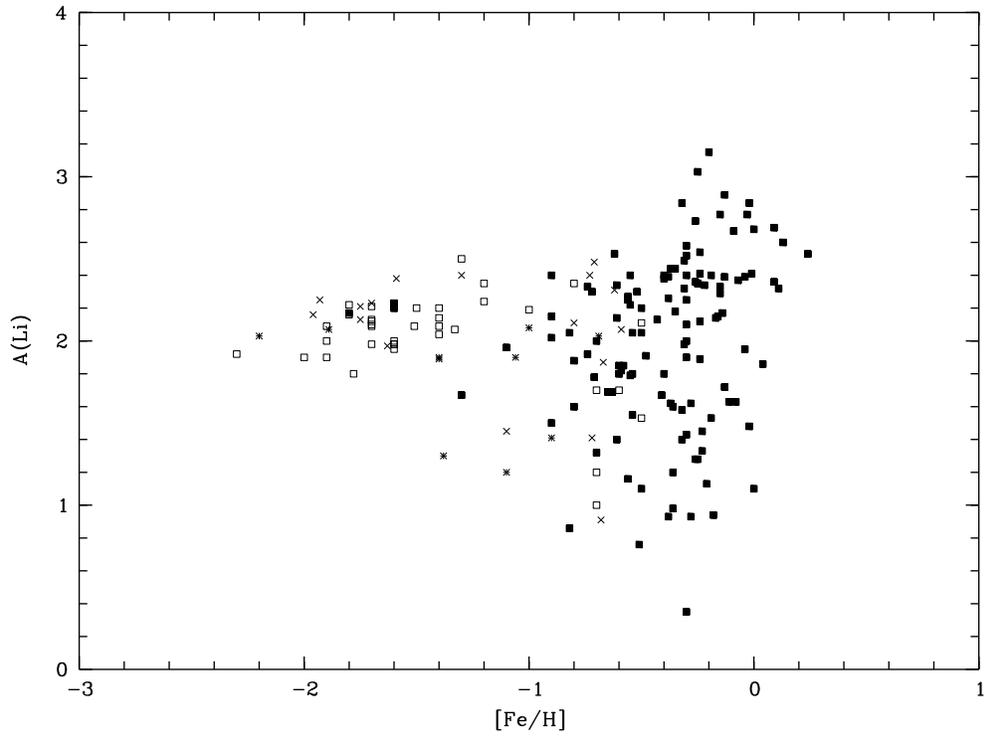,width=15cm,angle=-90}
\caption[]{\label{fig:fig 2} The observational A(Li) vs [Fe/H] diagram. 
Filled squares: disk stars; empty squares: non-disk stars; asterisks: objects 
with V$_{\mathrm{rad}}$ determination for which we were not able to provide an 
estimate on the kinematical membership; crosses: objects with no V$_{\mathrm
{rad}}$ determination. All the entries of Tab.4 have been plotted.}
\end{figure}
\twocolumn

\section{Stellar sites for $^7$Li production}

\subsection{Novae as lithium factories}

Both theoretical and observational evidence suggests that classical novae may 
be responsible for a non-negligible  contribution to the ISM pollution in 
several nuclides ($^7$Li, $^{13}$C, $^{15}$N, $^{17}$O) and radioactive 
isotopes ($^{22}$Na, $^{26}$Al) (see Gehrz et al.\,,1998 for a review on this 
subject). The nova explosion results from thermonuclear runaway on the surface 
of a CO or ONeMg white dwarf (WD) accreting hydrogen-rich matter from a main 
sequence companion which fills its Roche lobe in a close binary system; so the 
material ejected into the ISM shows a composition enriched in elements 
synthesized by explosive hydrogen-burning.\\
The pre-explosion $^3$He concentration in the accreted envelope plays a key 
role in determining how much $^7$Li will be produced in the outburst. 
Starrfield et al.\,(1978) found that the $^7$Li yield does scale linearly with 
the initial $^3$He content. Their result was used by D'Antona \& Matteucci 
(1991) who successfully reproduced the upper envelope of the observed A(Li) 
vs [Fe/H] diagram for the solar neighborhood. These authors adopted A(Li) = 
2.10 dex as the primordial lithium abundance and explained the observed 
increasing trend in lithium abundance with time as due to Galactic enrichment 
coming from lithium production by stellar sources, identified in both novae 
and AGB stars.\\
Later, novae nucleosynthesis was discussed by Boffin et al.\,(1993), who 
reinvestigated $^7$Li production in explosive hydrogen-burning with the 
inclusion of the $^8$B\,(p,$\gamma$)\,$^9$C reaction and ruled out novae 
as lithium producers at a Galactic level. In particular, they ruled out 
the linear dependence of Li production on the $^3$He abundance in the 
exploding envelope, which was a key ingredient in the D'Antona \& Matteucci 
model. 
Boffin et al. used a parameterized one-zone explosive nucleosynthesis model 
and detailed numerical network calculations to demonstrate how the high peak 
densities prevailing at the base of the hydrogen-burning shell during the 
nova outburst prevent the build-up of $^7$Li amounts sufficient for ISM 
enrichment. However, they also stressed the need for testing their results by 
detailed hydrodynamic nova models.\\
This check was effectively carried out a few years later: Hernanz et 
al.\,(1996), using a hydrodynamic code able to treat both the accretion 
and the explosion stages, have obtained large overproduction factors relative 
to the solar abundance for $^7$Be - and hence $^7$Li. Therefore, although the 
final masses injected in the ISM are small, significant production of $^7$Li 
by novae seems possible. In their models, these authors have assumed a solar 
composition of the infalling material and the existence of processes able 
to mix it with the inner layers of the underlying white dwarf. This assumption 
allows one to obtain the enhanced CNO or ONeMg abundances required in order to 
give rise to a nova outburst and to explain some observations (see Livio, 1994 
for a recent review).\\
More recently, Jos\'e \& Hernanz (1998) computed an enlarged grid of 
hydrodynamical nova models for both CO and ONeMg WDs, spanning a total 
range of white dwarf masses of 0.8\,--\,1.35 M$_\odot$. In these models 
different mixing levels - ranging from 25 to 75\% - between the accreted 
envelope and the underlying WD core were assumed. These models predict 
ejected quantities of $^7$Li able to affect the evolutionary history of 
this nuclide on a Galactic scale.\\
At this point, it is worth emphasizing the strong dependence of these results 
on the chemical composition at the onset of the explosion: the predicted 
$^7$Li overproduction factors will be correct only if the evolution of $^3$He 
during the accretion phase has been followed in the right manner. Moreover, we 
note that if the underlying WD is a CO one, the predicted $^7$Li abundances 
are $\sim$ 1 order of magnitude larger than in the case of an underlying 
ONeMg WD (see e.g.\,Table 1 of Hernanz et al.\,, 1996).

\subsection{Lithium production in massive AGB stars}

$^7$Li production in AGB stars is the only Li stellar production mechanism 
supported by observations.

Smith \& Lambert (1990) analysed high-resolution spectra for 27 red giants in 
the Magellanic Clouds, spanning a range in bolometric absolute magnitudes 
M$_{bol}$ $\sim$ $-$\,5 to $-$\,9. They found that Li-rich stars were confined 
to those with M$_{bol}$ $\sim$ $-$\,6 to $-$\,7, while lower (M$_{bol}$ $\sim$ 
$-$\,5 to $-$\,5.5) and higher (M$_{bol}$ $\sim$ $-$\,7 to $-$\,9) luminosity 
red giants showed no Li features. In addition, all their Li-strong stars presented C/O $<$ 1 and bore evidence of strengthened s-process atomic lines. These 
facts were interpreted as a signature of the dredge-up and subsequent envelope 
burning mechanism, occuring in the massive AGB stars (M $\sim$ 4\,--\,8 
M$_\odot$). Since every luminous AGB star observed showed the Li\,I doublet, 
it seemed to be very likely that the $^7$Li produced in the outer envelope was 
not destroyed quickly, but it was surviving in the stellar atmosphere and 
injected into the ISM by stellar winds. Smith \& Lambert derived a maximum 
abundance of $^7$Li as large as A(Li) $\sim$ 4.0 for the luminous AGB stars 
and indicated them as a major source of $^7$Li in a galaxy.\\
These observational findings were confirmed by theoretical calculations in 
which a time-dependent ``convective diffusion'' algorithm for the hot bottom 
envelopes of AGB stars was coupled with a fully self-consistent evolutionary 
sequence (Sackmann \& Boothroyd, 1992). It was shown that values of A(Li) 
lying between 4 and 4.5 could be obtained for stars in the luminosity range M~
$_{bol}$ $\sim$ $-$\,6 to $-$\,7, in excellent agreement with observations of 
the most lithium-rich giants in external galaxies (Smith \& Lambert, 1990; 
Smith et al.\,, 1995) but in poorer agreement with observations in our Galaxy 
(Abia et al.\,, 1991). According to Sackmann \& Boothroyd, super-rich lithium
 giants would occur in the mass range M $\sim$ 4\,--\,6 M$_\odot$, when 
the temperature at the base of the convective envelope exceeds 50 $\times$ 
10$\,^6$~ K and the Cameron-Fowler mechanism\footnote{See Cameron \& Fowler 
(1971) for a detailed description of this mechanism.} works (in these models, 
stars with M $\ge$ 7 M$_\odot$ ignite carbon in the center before they become 
AGB stars and never experience hot bottom burning).\\
Later, Plez et al.\,(1993) enlarged the Smith \& Lambert's sample of giants 
in the Small Magellanic Cloud (SMC) and pointed out that several of these 
metal-poor giants have atmospheric abundances of $^7$Li too low to provide a 
significant contribution to Galactic enrichment. On the other hand, Sackmann 
\& Boothroyd had shown that a decrease in the metallicity causes the hot 
bottom burning to start at smaller masses, resulting in a larger $^7$Li 
production! This seems to lead to a discrepancy with Plez et al.'s 
observations, but it should be recalled that low metallicity giants lose their 
envelope mass at a smaller rate than more metal-rich giants, thus allowing 
Li-burning to occur before the planetary nebula (PN) ejection and thus 
preventing significant ISM pollution and explaining Plez et al.'s findings.\\
Recent studies (Mazzitelli et al.\,, 1999) confirm that $^7$Li abundances as 
large as A(Li) $\sim$ 4.0 dex at maximum can be found in the atmospheres of 
massive AGB stars.

The $^7$Li production by C-stars (coming from progenitors in the mass range 
2\,--\,5 M$_\odot$) can be regarded as negligible on a Galactic scale, on the 
ground of the very small number of Li-rich C-stars detected with respect to 
their total number (see Wallerstein \& Conti, 1969; Abia et al.\,, 1993a, 
1993b; but see also Beers et al.\,, 1992). Such a low number of Li-rich 
C-stars can be easily explained if it is assumed that these stars represent a 
short-lived phase with respect to the overall stellar lifetime on the AGB. On 
the contrary, one should assume that for stars with progenitor masses in the 
range 5\,--\,8 M$_\odot$ the envelope ejection in the PN stage shortly 
follows the lithium production in the envelope itself. We want to recall 
briefly that lithium production in these stars is thought to be due either to 
hot bottom burning, occuring during thermal pulses (TPs) at the base of the 
common envelope, or to a mechanism such as that described by Iben (1973), 
involving only the region of the outer convective envelope lying near the 
hydrogen burning shell, during the long timescale of the interpulse phases.

\subsection{$^7$Li synthesis in Type II supernova explosions}

$^7$Li synthesis in massive stars (M $>$ 10 M$_\odot$) is theoretically 
explained as being due to a particular mechanism, the neutrino process. 
The first realistic exploration of the so-called neutrino process 
($\nu$-process), acting during SNeII explosion, was undertaken by Woosley et 
al.\,(1990). Such a process occurs in the shells overlying the collapsing core 
of a contracting massive star. In these conditions, the flux of neutrinos is 
so large that, despite the small cross section, evaporation of neutrons or 
protons from heavy elements and helium is expected. The back reaction of these 
nucleons on other species alters the outcome of traditional nucleosynthesis 
calculations, resulting in a large production of a great number of rare 
isotopes. $^7$Li is one of these; it is thought to be made mainly in the 
helium and in the silicon shells from $\mu$- and $\tau$-neutrinos interacting 
with helium. In their work, Woosley et al. strongly supported the idea that 
this lithium production by $\nu$-process in massive stars could be large 
enough to explain the full $^7$Li Solar System abundance.\\
This point was later revised when Timmes et al.\,(1995), using the output 
from a grid of 60 Type II supernova models of varying mass and metallicity 
(Woosley \& Weaver, 1995), computed the chemical evolution of several stable 
isotopes, taking into account also the nucleosynthesis from Type Ia supernovae 
and from single stars with M $\le$ 8 M$_\odot$. They found that massive stars 
are producing lithium prior to [Fe/H] $\sim$ $-$\,1.0 dex, but until this 
metallicity value the contributions are small compared to the infall values, 
thus preserving the flat shape of the diagram A(Li) vs [Fe/H] inferred 
from the observations. Finally, they concluded that Type II supernovae 
contribute about one-half the solar $^7$Li abundance\footnote{This result is 
achieved for $\mu$- and $\tau$-neutrino temperatures in the range 6\,--\,8 
MeV, which is the range suggested by SN\,1987\,A.}, pointing to a lower $^7$Li 
production rate by the $\nu$-mechanism than Woosley et al.\,(1990). Such an 
outcome was confirmed also by Matteucci et al.'s (1995) analysis, where 
C-stars plus massive AGB stars on the one hand and Type II SNe on the other 
hand were found to contribute each one-half the Solar System $^7$Li abundance, 
although the authors did not conclusively rule out a fraction between 1/4 and 
3/4 from both sources, because of the uncertainties in the input 
nucleosynthesis.

\section{The model}

\subsection{Basic assumptions}

The adopted model of chemical evolution is that of Chiappini et al.\,(1997), 
in which we included the $^7$Li evolution, taking into account Li production 
from all the stellar sources identified above and from GCR spallation. This 
model assumes that the halo and thick-disk formed quickly (on a timescale of 
$\sim$ 2 Gyr) during a first infall episode and the thin-disk formed on a much 
larger timescale ($\sim$ 8 Gyr) during a second independent infall episode.

The nova system nucleosynthesis has been included in the model in a detailed 
way under simple hypotheses. We first computed the nova systems formation rate 
at the time $t$ as a fraction of the formation rate of white dwarfs at a 
previous time $t\,-\,\Delta\,t$ as in D'Antona \& Matteucci (1991):
\begin{displaymath}
R_{novae}(t) = \alpha\,\int_{0.8}^{8} \psi(t\,-\,\tau_{m}\,-\,\Delta\,t)\,
\phi(m)\,dm\,.
\end{displaymath}
Here $\Delta\,t$ is a delay time whose value has to be fixed to guarantee the 
cooling of the WD at a level that ensures a strong enough nova outburst. 
We chose $\Delta\,t$\,=\,1 Gyr as a suitable average value (D'Antona, 1998) 
and assumed that all stars with initial masses in the range 0.8\,--\,8 
M$_\odot$ end their lives as WDs.\\
$\psi(t)$ is the star formation rate (SFR), $\tau_{m}$ is the lifetime of the 
star of mass $m$ and $\phi(m)$ is the initial mass function (IMF). More about 
SFR and IMF parameterization can be found in Chiappini et al.\,(1997) 
concerning the disk of the Galaxy and in Matteucci et al.\,(1999) concerning 
the Galactic bulge.\\
The rate of nova eruptions is related to the WD formation rate:
\begin{displaymath}
R_{outbursts} = \alpha\,R_{WDs}\,n,
\end{displaymath}
where $\alpha\,R_{WDs}$ is the formation rate of WDs in binary systems which 
will give rise to nova eruptions, and $n$ = 10$\,^{4}$ is the average number 
of nova outbursts for a typical nova system (Bath \& Shaviv, 1978; see also 
Shara et al.\,, 1986).\\
The parameter $\alpha$, set equal to a constant value along the overall 
evolutionary history of the Galaxy, can be fixed by the rate of nova outbursts 
in our galaxy at the present time. Estimates of this quantity in the current 
literature range from as few as 11 to as many as 260 nova outbursts 
yr$\,^{-1}$. In particular, predictions based on scalings from extragalactic 
nova surveys suggest low values (11\,--\,46 yr$\,^{-1}$, Ciardullo et al.\,, 
1990; 15\,--\,50 yr$\,^{-1}$ - with the lowest values, between 15 and 25 
yr$\,^{-1}$, strongly favored - Della Valle \& Livio, 1994), whereas estimates 
based on extrapolations of Galactic nova observations give the highest rates 
(73 $\pm$ 24 yr$\,^{-1}$, Liller \& Mayer, 1987; 260 yr$\,^{-1}$, Sharov, 
1972; 50 yr$\,^{-1}$, Kopylov, 1955; 100 yr$\,^{-1}$, Allen, 1954)
\footnote{Hatano et al.\,1997 reanalyzed Liller \& Mayer's data and argued 
that the correct rate should be $\sim$ 41 yr$\,^{-1}$ rather than 73 $\pm$ 24 
yr$\,^{-1}$.}. We chose for the present time rate of nova outbursts in the 
Galaxy $R_{outbursts}(t_{Gal})$ $\sim$ 25 yr$\,^{-1}$ for the following 
reasons: (1) observation of novae in nearby galaxies would avoid, or at least 
minimize, most of the problems encountered by Galactic observations; (2) a 
recent study of Shafter (1997) shows that the nova rate based on Galactic 
observations can be made consistent with the rate predicted from the 
extragalactic data, particularly if the Galaxy has a strong bar oriented in 
the direction of the Sun (in this latter, most favourable case, the suggested 
value is near $\sim$ 20 yr$\,^{-1}$, otherwise, if the bar is weak or 
misaligned, the global rate can be reduced only to $\sim$ 30 yr$\,^{-1}$). 

\subsection{Nucleosynthesis prescriptions}

We assumed a homogeneus Big Bang $^7$Li abundance of 2.2 dex (Bonifacio \& 
Molaro, 1997) as the primordial one and considered all the contributions from 
the various classes of stellar $^7$Li factories seen in \S 3 in the Galactic 
chemical evolution model.

In this study we adopted the updated lithium yields from theoretical nova 
outbursts provided by Jos\'e \& Hernanz (1998). We took the mean mass ejected 
as $^7$Li averaged on 7 evolutionary sequences for CO WDs and the mean mass 
ejected as $^7$Li averaged on so many evolutionary sequences for ONeMg WDs. 
ONeMg WDs are believed to originate from stars with initial masses in the 
range 7 M$_\odot$\,--\,M$_{up}$\footnote{Here we assume M$_{up}$, the limiting 
mass for the formation of a degenerate CO core, equal to 8 M$_\odot$, although 
some authors suggest 5\,--\,6 M$_\odot$ as a more suitable value, when 
overshooting is taken into account (Chiosi et al.\,, 1992; Marigo et al.\,, 
1996).} (but there is still debate on this point). Since the lifetime of a 7 
M$_\odot$ star is $\sim$ 0.045 Gyr, we assumed that for $t$ $\le$ 0.045 Gyr 
only ONeMg WDs can contribute to nova systems; for times larger than 0.045 
Gyr, about 30\% of novae occur in systems containing ONeMg WDs, while the 
remaining take place in systems accreting hydrogen-rich envelopes onto CO 
WDs.\\
In particular, the prescriptions for $^7$Li yields from novae we used in our 
model are as follows:\\
for $t$ $<$ 0.045 Gyr we assumed
\begin{center}
M$_{ej}$ = 1.95 $\times$ 10$\,^{-1}$ M$_\odot$
\end{center}
(mean mass ejected by a single nova system during its overall evolution) and 
\begin{center}
X$_{^7Li}$ = 9.24 $\times$ 10$\,^{-7}$ 
\end{center}
(mean $^7$Li abundance - in mass fraction - in the ejected material).\\
For $t$ $>$ 0.045 Gyr we assumed 
\begin{center}
M$_{ej}$ = 2.63 $\times$ 10$^{-1}$ M$_\odot$
\end{center}
with 
\begin{center}
X$_{^7Li}$ = 2.85 $\times$ 10$\,^{-6}$ 
\end{center}
as the mean lithium abundance in the ejected envelope.

As far as AGB stars are concerned, we included here a metallicity dependence 
of the lithium yields from massive AGB stars as in Matteucci et al.\,(1995), 
which accounts for the lower $^7$Li abundances observed in the low metallicity 
SMC AGB stars relative to the higher ones exhibited by their more metal-rich 
Galactic counterparts. We assumed no production of lithium by AGB stars until 
a metallicity of Z = 10$\,^{-3}$, then allowing lithium production only in the 
mass range 5 M$_\odot$\,--\,M$_{up}$, where M$_{up}$ increases from 5 to 8 
M$_\odot$ as Z reaches the solar value. The $^7$Li abundance in the ejected 
material is assumed to be either A(Li) = 4.15 or A(Li) = 3.50 dex (see Tab.1 
for different model prescriptions). Since only a small number of Li-rich 
C-stars (coming from progenitors with masses in the range 2\,--\,5 M$_\odot$) 
are known out of hundreds of C-stars observed in the Galaxy and in the 
Magellanic Clouds, we assumed their contribution to the lithium enrichment to 
be almost negligible and followed the D'Antona \& Matteucci (1991) 
prescriptions. In one model (model C) we also completely suppressed this class 
of lithium factory.

To account for lithium production in Type II supernovae, we considered the 
metallicity dependent $^7$Li yields given by Woosley \& Weaver (1995). We 
included in our computation both the full yields (model A and B) and those 
reduced to a half (model C).

$^7$Li astration in stars of all masses has also been taken into account.

The contribution of GCR spallation has been taken into account by 
incorporating the absolute yields from Lemoine et al.\,(1998) into the 
chemical evolution model. We took the yields from the lower-bound spectrum 
in their Table 1.

\begin{table}
\centering
\caption[]{\label{tab:tab 1} Nucleosynthesis prescriptions.}
\paragraph{}
\begin{tabular}{c c c c c} 
\multicolumn{5}{c}{}\\
\hline
Model&C-stars&M-AGB&SNeII&novae\\
\hline
A&A(Li) = 3.85&A(Li) = 4.15&WW95&no\\
B&A(Li) = 3.85&A(Li) = 4.15&WW95&yes\\
C&no&A(Li) = 3.50&WW95/2&yes\\
\hline
\end{tabular}
\end{table}

\section{Results}

In Fig.3 we show the predicted log\,($R_{WDs}$), log\,($R_{novae}$) trends vs 
time in the solar neighborhood for models with $\alpha$ = 0.0125. By assuming 
a Galactic scale height of 300 pc for WDs - and hence novae - and a Galactic 
volume of 10$\,^{11}$ pc$\,^{3}$ we obtain $R_{outbursts}(t_{Gal})$ = 24.5 
yr$\,^{-1}$, in good agreement with observations (see section 4.1).\\
We predict also that the current WD birthrate in the Galaxy should be 
$R_{WDs}(t_{Gal})$ = 2.27 $\times$ 10$\,^{-12}$ WDs pc$\,^{-3}$ yr$\,^{-1}$, 
to be compared with observational estimates which give $R_{WDs}(t_{Gal})$~ = 
1 $\times$ 10$\,^{-12}$ WDs pc$\,^{-3}$ yr$\,^{-1}$ (Yuan, 1989) and $D_{WDs}$ 
(spatial density) $\sim$ 1 $\times$ 10$\,^{-2}$ pc$\,^{-3}$ (Weidemann, 1967). 
This slight discrepancy, however, can be explained by recent dynamical 
calculations (Chamcham, 1998) suggesting that about 40\% of the white dwarfs 
originally formed in the thick-disk have moved from their birthplaces so that 
they should be observed at a higher scale height at the present time.

\begin{figure}
\psfig{figure=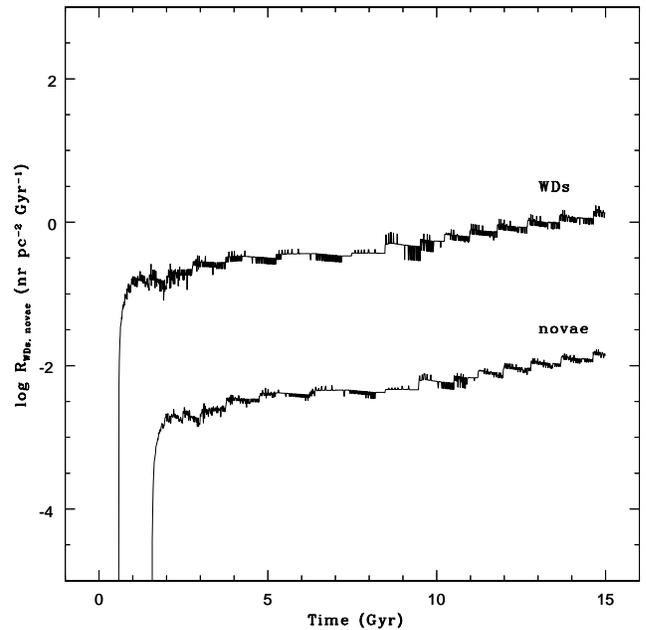,width=9.1cm,angle=0}
\caption[]{\label{fig:fig 3} Birthrates of white dwarfs and nova systems as 
functions of time, predicted with the two-infall Galactic chemical evolution 
model that considers a threshold in the surface gas density below which star 
formation is suppressed (see text for details). The nova systems formation 
rate is assumed to be a fraction $\alpha$~ = 0.0125 of the WDs formation rate.}
\end{figure}

\begin{figure}
\psfig{figure=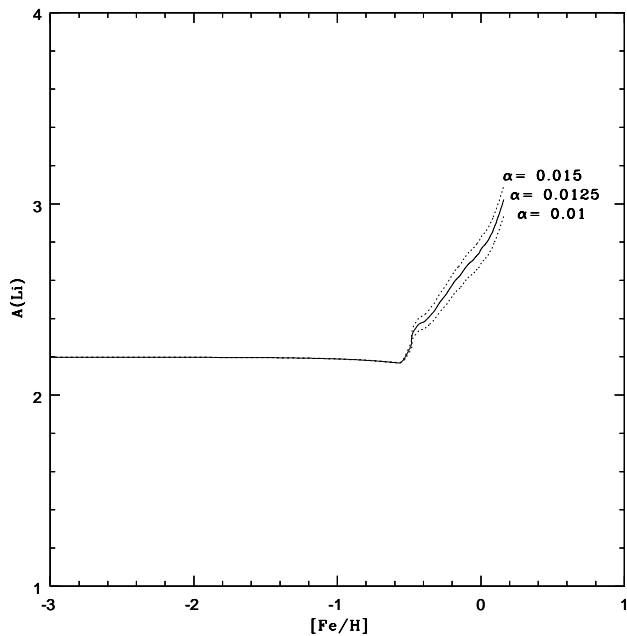,width=9.1cm,angle=0}
\caption[]{\label{fig:fig 4} A(Li) vs [Fe/H] theoretical trends predicted for 
the solar neighborhood by three models taking into account lithium production 
in nova outbursts under slightly different assumptions on the percentage of 
newly formed WDs that enters in the building-up of nova systems.}
\end{figure}

Note that the oscillations in the theoretical curves in Fig.3 are caused 
by the introduction of a threshold in the surface gas density (7 M$_\odot$ 
pc$\,^{-2}$) below which star formation stops, owing to gas instability 
against density condensations in these conditions (see Chiappini et al.\,, 
1997). The presence of such a threshold leads also to a delay in the WD 
formation, which starts only after $\sim$ 0.6 Gyr (compare Fig.3 here with 
Fig.2 in D'Antona \& Matteucci, 1991, whose model does not include such a 
threshold in the gas density).

\begin{figure}
\psfig{figure=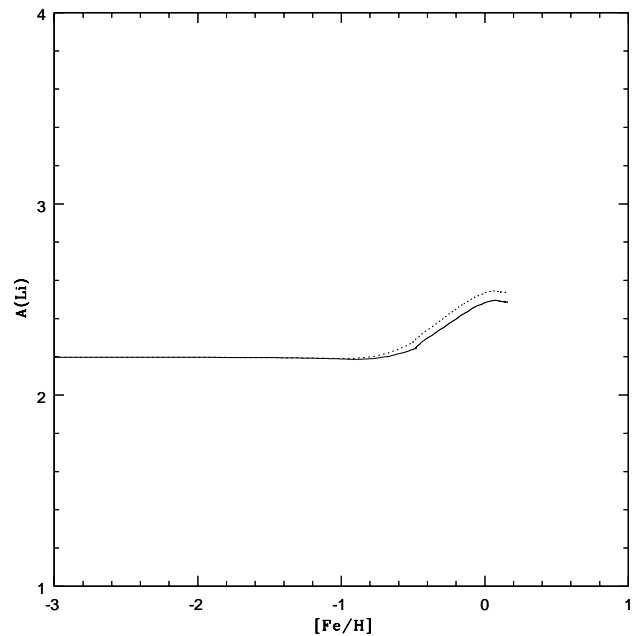,width=9.1cm,angle=0}
\caption[]{\label{fig:fig 5} Accounting for lithium production by both 
C-stars and massive AGB stars (dotted line) one obtains a slightly higher 
$^7$Li abundance at the present time than in the case in which only massive 
AGB stars are considered (continuous line).}
\end{figure}

We ran several models considering separately all the $^7$Li stellar sources 
discussed above. The Galactic lithium enrichment in the solar neighborhood 
due only to novae is sketched in Fig.4: novae start injecting material into 
the ISM with a time delay of $\sim$ 2 Gyr, when the ISM has already achieved a 
metallicity [Fe/H] $\sim$ $-$\,0.5 dex. This long time-lag in the occurrence 
of an ISM pollution by novae is the direct consequence of two elements 
characterizing the evolution of such systems, both acting in the same 
direction: 1) the time-lag required to form the WD and 2) the time necessary 
for the WD to cool enough to allow strong nova outbursts.\\
The effects produced by changing $\alpha$ from 0.0125 to smaller or greater 
values on the predicted A(Li) vs [Fe/H] trend are also shown in Fig.4. We 
analysed two possible different choices, $\alpha$ = 0.01 and $\alpha$ = 0.015. 
The first choice leads to $R_{outbursts}(t_{Gal})$ = 19.6 yr$\,^{-1}$, the 
second one to $R_{outbursts}(t_{Gal})$ = 29.4 yr$\,^{-1}$. These estimates 
reproduce very well the lower and upper limits of the permitted values 
inferred from the analysis of Shafter (1997). We have chosen $\alpha$ = 0.0125 
as the most suitable value and used it in all the calculations.

In Fig.5 we show the evolution of the $^7$Li abundance in the solar 
neighborhood as predicted by two models both allowing lithium production to 
happen only in AGB stars (prescriptions on lithium production as in model A). 
If we assume $^7$Li production to take place in both C-stars and massive AGB 
stars (dotted line) we obtain a slightly higher $^7$Li abundance than assuming 
$^7$Li production only in massive AGB stars (continuous line). 

\begin{figure}
\psfig{figure=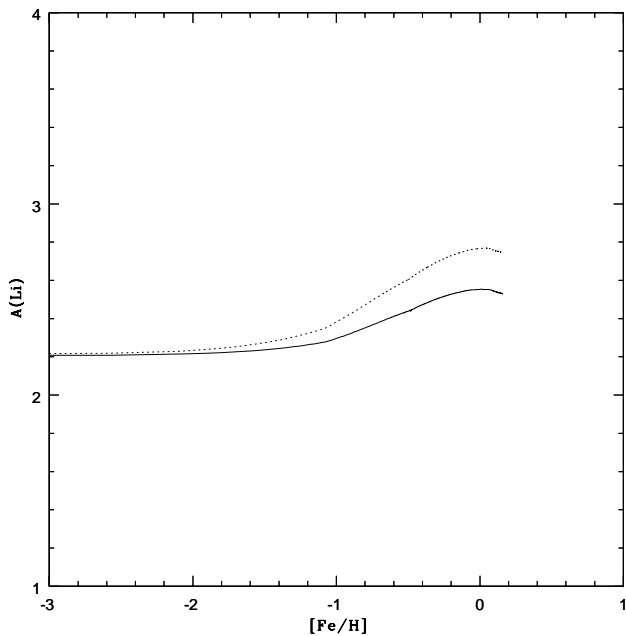,width=9.1cm,angle=0}
\caption[]{\label{fig:fig 6} Theoretical predictions for the trend A(Li) vs 
[Fe/H] from two models when only $^7$Li production from massive (M $>$ 10 
M$_\odot$) stars is allowed. Dotted line: we used the Woosley \& Weaver (1995) 
$^7$Li\,+\,$^7$Be yields; continuous line: we used the same yields but reduced 
to a half. Note that the metallicity dependence of such yields guarantees the 
flatness of the Spite plateau until [Fe/H] $\sim$ $-$\,1.0.}
\end{figure}

In Fig.6 we show the effect of $^7$Li production only by massive stars. We 
used either the full $^7$Li\,+\,$^7$Be yields (Woosley \& Weaver, 1995) or 
the same yields reduced to one half.

From figures 4, 5 and 6 we can immediately see how novae, giving rise to a 
strong Li-enrichment at high metallicities, can in principle account for the 
present $^7$Li abundance in the gas without any other stellar source. However, 
acting on evolutionary timescales as long as 1.5 Gyr at least, they cannot 
justify at the same time the rise off the Spite plateau. On the other hand, 
Type II supernovae start restoring their $^7$Li into the ISM at earlier times, 
but they are not able to fully explain the present gas content of lithium, if 
they are not coupled to other stellar Li producers. The AGB stars present a 
quite similar behaviour, but restore their $^7$Li a bit later than SNeII, and 
in minor (if not almost equal) amounts (compare Figs.5 and 6).\\
Therefore, one single stellar category of $^7$Li producers could never 
explain all the observed features of the diagram A(Li) vs [Fe/H], under 
realistic prescriptions about $^7$Li synthesis!

Therefore we computed three different chemical evolution models for both the 
solar vicinity and the Galactic bulge, adding together the different stellar 
sources.

\begin{figure}
\psfig{figure=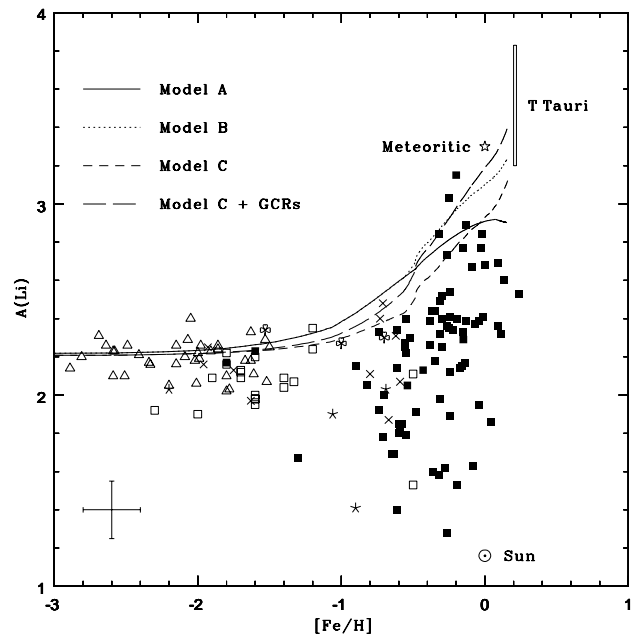,width=9.1cm,angle=0}
\caption[]{\label{fig:fig 7} A(Li) versus [Fe/H] theoretical predictions for 
the solar neighborhood from models A, B, C, and C + GCRs (see text), compared 
with the observational diagram coming out from our data analysis. The upper 
limits have been removed from the sample. Filled symbols: disk stars; empty 
symbols: non-disk stars (triangles: halo stars from Bonifacio \& Molaro, 1997; 
squares: stars from our data-base). Crosses and asterisks identify stars 
without kinematical membership determination; clovers are objects with 
multiple $^7$Li determination in the literature for which we took the average 
value. Solar, meteoritic (Anders \& Grevesse, 1989) and T\,Tauri (see text for 
references) $^7$Li abundances are also shown.}
\end{figure}

\begin{figure}
\psfig{figure=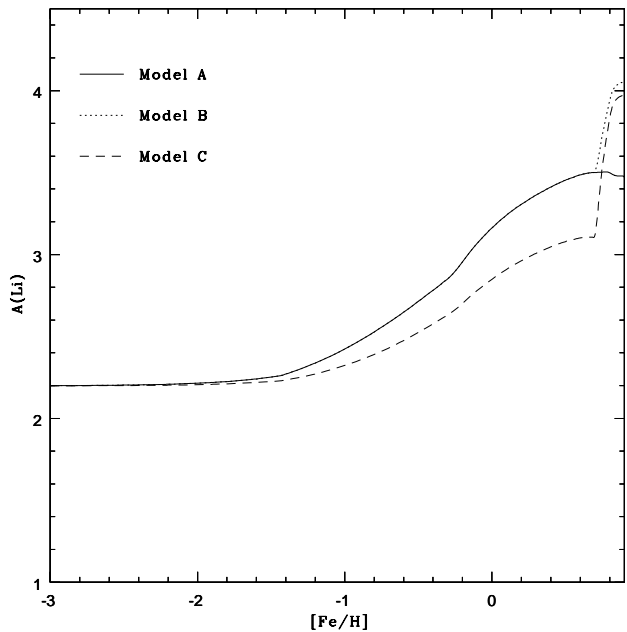,width=9.1cm,angle=0}
\caption[]{\label{fig:fig 8} Theoretical predictions on $^7$Li abundance 
evolution in the ISM in the Galactic bulge. Model A (best model in Matteucci 
et al.\,, 1999), accounting for $^7$Li production by all stellar sources but 
novae, predicts the lowest present lithium abundance in the gas in the central 
region.}
\end{figure}

In Figs.7 and 8 we sketch the A(Li) vs [Fe/H] trends obtained in the case of 
the solar neighborhood evolution and in the case of the bulge evolution, 
respectively. Models A, B and C are referring to the different nucleosynthesis 
prescriptions outlined in Tab.1. The observational points plotted in Fig.7 are 
those given in Tab.4 after the upper limits have been removed. An average 
value has been taken for the three objects with multiple $^7$Li detections 
which are tracing the upper envelope (see the appendix). From Fig.7, we see 
that model A, using the nucleosynthesis prescriptions from the best model of 
Matteucci et al.\,(1995), does not reproduce the highest values of A(Li), 
found in some of the most metal-rich stars, nor the observed upper envelope in 
the metallicity range $-$\,1.5 $\le$ [Fe/H] $\le$ $-$\,0.5 dex. Model B, 
constructed by adding the contribution from nova outbursts to the previous 
one, produces a curve which becomes much steeper at the highest metallicities, 
hardly reaching the highest Li abundances exhibited by the most metal-rich 
stars in our sample. In model C, lithium production from C-stars is set to 
zero, and $^7$Li yields from both massive AGB stars and TypeII SNe are 
reduced. In particular, SNeII yields are assumed to be one half of the 
original ones. As one can see from Fig.7, model C reproduces at best the 
observational data in the metallicity range $-$\,1.5 $\le$ [Fe/H] $\le$ 
$-$\,0.5, but fails in reproducing the meteoritic and the T\,Tauri $^7$Li 
abundances. $^7$Li production from GCR spallation could be helpful in 
reproducing the highest observed $^7$Li abundances.

Molaro et al.\,(1997), by studying a large sample of stars with $^9$Be 
determinations, found that A(Be)\footnote{A(Be) = log$_{10}$(N$_{^9Be}$/N$_H$) 
+ 12.} correlates linearly with [Fe/H], with a slope confirmed also by Duncan 
et al.\,(1997) and Garc\'\i a-L\'opez et al.\,(1998). $^9$Be is produced only 
by GCR spallation. Using the Steigman \& Walker (1992) formula,  $^7$Li/$^9$Be 
= 7.6 for PopII stars with T$_{\mathrm{eff}}$ $<$ 6200 K, one can see that the 
$^7$Li amount expected from GCR spallation alone is about 1\% of the 
primordial $^7$Li abundance at [Fe/H] $\sim$ $-$\,2.5, and that it becomes 
more important with increasing metallicity, being about 25\% around [Fe/H] 
$\sim$ $-$\,1.0. At larger metallicities one could expect an even larger 
contribution, although other authors suggest that GCRs can contribute to the 
present ISM $^7$Li abundance by no more than $\sim$ 10\%, on the basis of the 
$^7$Li/$^6$Li ratio towards $\zeta$ Oph (e.g. Lemoine et al.\,, 1995). 
In order to include GCR nucleosynthesis in our chemical evolution model in a 
self-consistent way, we used the absolute yields provided by Lemoine et al.\,
(1998) for various metallicities. As a result, we produced a smoother rise 
from the plateau, and we were finally able to match the meteoritic and 
T\,Tauri lithium abundances (see Tab.2).\\
The lithium abundance in the Galactic bulge at the present time predicted by 
our model C is as high as $\sim$ 4 dex (Fig.8); if novae are not included in 
the computation, a lower present Li abundance is expected in the bulge ($\sim$ 
3.5 dex, see also Matteucci et al.\,, 1999). 
However, this is only a lower limit, since $^7$Li production from GCRs was not 
included in the bulge model. In fact, giving the uncertainties in the relevant 
parameters (see Lemoine et al.\,, 1998), it seems not so meaningful to 
translate the GCR nucleosynthesis results for the solar neighborhood to the 
central region of the Galaxy.
\begin{table}
\centering
\caption[]{\label{tab:tab 2} $^7$Li abundances at the epoch of Solar System 
formation and in the interstellar medium as predicted by our four models 
for the solar neighborhood.}
\paragraph{}
\begin{tabular}{c c c} 
\multicolumn{3}{c}{}\\
\hline
Model&A(Li)$_{\mathrm{SS}}$&A(Li)$_{\mathrm{ISM}}$\\
\hline
A&2.91&2.90\\
B&3.12&3.24\\
C&2.95&3.12\\
C + GCRs&3.21&3.39\\
\hline
\end{tabular}
\end{table}

\section{Conclusions}

In this paper we have computed the evolution of the $^7$Li abundance in the 
ISM of the solar neighborhood and the bulge of our Galaxy.\\
We took into account several stellar $^7$Li factories (novae, massive AGB 
stars, C-stars and Type II SNe), together with the GCR contribution. In 
particular, we adopted new nucleosynthesis prescriptions for novae (Jos\'e 
\& Hernanz, 1998) and the yields of $^7$Li of Lemoine et al.\,(1998) as far 
as GCR nucleosynthesis is concerned.\\
We  compiled a new data sample for $^7$Li abundances in the solar neighborhood 
stars identifying the stars belonging to disk and halo Galactic components on 
the basis of their kinematics. The identification of metal-rich Li depleted 
halo stars provides evidence that Li depletion depends from metallicity only, 
thus supporting the use of the upper envelope in the Li data-set as the true 
indicator of Li Galactic evolution. The selection of the stars which have 
likely preserved their $^7$Li content reveals a possible extension of the Li 
plateau towards higher metallicities, up to [Fe/H] $\sim$ $-$\,0.5, $-$\,0.3, 
with a steeper rise afterwards.\\
Comparing theoretical predictions with the data we derived the following 
conclusions:
\begin{itemize}
\item[1\,--] In order to reproduce the upper envelope of the A(Li) vs [Fe/H] 
diagram we need to take into account several stellar Li sources: AGB stars, 
Type II SNe and novae. In particular, novae are required to reproduce the 
steep rise of A(Li) between the formation of the Solar System and the present 
time, as is evident from the data we sampled. On the other hand, $^7$Li 
yields for SNeII should be lowered by at least a factor of two in order to 
reproduce the extension of the Spite plateau.
\item[2\,--] We produced arguments suggesting that GCRs could be responsible 
for the production of a non-negligible amount of Li at metallicities larger 
than [Fe/H] $\sim$ $-$\,1.0 dex. In particular, we showed that without any GCR 
contribution it is impossible to reach the high values observed in meteorites 
and T\,Tauri stars: GCRs are responsible for $\sim$ 45\% of the Solar System 
$^7$Li (prescriptions on stellar nucleosynthesis as in model C).
\item[3\,--] By adopting the nucleosynthesis prescriptions of our model C 
for the Galactic bulge, we predicted a lower limit for the present time $^7$Li 
abundance in this central region of the Galaxy of the order of 4.0 dex, which 
is 0.5 dex higher than previously estimated in Matteucci et al.\,(1999) 
neglecting the novae contribution.
\end{itemize}

If new $^7$Li measurements in stars with [Fe/H] around $-$\,0.5 dex will 
confirm the extension of the plateau towards such high metallicities, a 
revision of the contribution to the $^7$Li abundance from GCR spallation 
too could be needed.

\section*{Appendix}

\subsection*{Star by star analysis}

In order to identify the stars which can be actually considered as tracers of 
the upper envelope at every metallicity, and discard the depleted ones, we 
analysed in detail the abundances of all the stars defining the upper envelope 
in the A(Li)\,--\,[Fe/H] diagram for metallicities larger than $\sim$ $-$\,1.5 
dex. The stars are listed below in order of metallicity.
\begin{itemize}
\item[1\,--] HIP\,42887: Deliyannis et al.\,(1990) give A(Li) $\le$ 2.50 and 
[Fe/H] = $-$\,1.30; Glaspey et al.\,(1994) reduce this upper limit to A(Li) 
$<$ 1.20 and assume [Fe/H] = $-$\,1.22. Since these are only upper limits for 
the Li abundance in this star, we do not take this object into account as a 
tracer of the upper envelope.
\item[2\,--] HIP\,99423: we assume A(Li) = 2.34 and [Fe/H] = $-$\,1.53 for 
this star (averaging on three measures, see Tab.4).
\item[3\,--] HIP\,3026: only one Li detection (A(Li) = 2.35, [Fe/H] = 
$-$\,1.20).
\item[4\,--] HIP\,86321: two A(Li) determinations in substantial agreement 
(from the compilation of Deliyannis et al.\,, 1990). We took their average 
value: A(Li) = 2.27. For the metallicity we preferred the value $-$\,1.00, 
according with the estimate in Cayrel de Strobel et al.\,(1992).
\item[5\,--] HIP\,108490: we listed three $^7$Li determinations for this star 
in Tab.4. They differ by 0.1\,--\,0.2 dex, the mean value being 2.30. This 
estimate agrees with the most recent determination of the atmospheric lithium 
abundance in this star by Stephens et al.\,(1997) (A(Li) = 2.39).
\item[6\,--] HIP\,79720 (A(Li) = 2.40, [Fe/H] = $-$\,0.73) and HIP\,29001 
(A(Li) = 2.48, [Fe/H] = $-$\,0.71). For these two crucial objects we found a 
Li detection only by  Lambert et al.\,(1991), and an independent confirmation 
is desirable considering their importance in the economy of the discussion.
\item[7\,--] The Li\,I feature in the atmosphere of HIP\,112935 is given as 
W(Li) = 47 m\AA\, by Deliyannis et al.\,(1990), referring to the detection 
by Duncan (1981), whereas Balachandran (1990) gives less than 1 m\AA\, (A(Li) 
$<$ 0.82). Moreover, both the Lambert et al.\,(1991) and Boesgaard et 
al.\,(1998) estimates (A(Li) $<$ 1.28 and A(Li) $\le$ 0.90, respectively) 
agree with the Balachandran one. These upper limits and the detection by 
Duncan are not consistent; we choose the upper limit as the 
correct indicator of the $^7$Li content in HIP\,112935.
\item[8\,--] HIP\,14181 (A(Li) = 2.31 and [Fe/H] = $-$\,0.62) and HIP\,21167 
(A(Li) = 2.34 and [Fe/H] = $-$\,0.61), both single detections from Lambert et 
al.\,(1991).
\item[9\,--] At [Fe/H] $\sim$ $-$\,0.5 there are some stars lying around A(Li) 
$\sim$ 2.40, then a steep rise occurs at [Fe/H] $\sim$ $-$\,0.3: HIP\,10306, 
HIP\,11783 and HIP\,46853 show that the ISM Li abundance rapidly increases to 
$\sim$ 3.2\,--3.3.
\end{itemize}

\begin{acknowledgements}
It is a pleasure to thank F. D'Antona for useful discussions about nova 
systems and AGB stars evolution, M. Della Valle for explanations concerning 
the observed nova outburst rate in the Galaxy at the present time and J. 
Danziger for useful comments during the course of this work. We also thank E. 
Casuso and J. Beckman for providing us with a copy of their paper prior to 
publication. Finally, we would like to thank an anonymous referee for his/her 
useful suggestions.\\
This research has made use of the SIMBAD data-base, operated at CDS, 
Strasbourg, France, of the HIPPARCOS catalogue and of NASA's Astrophysics 
Data System Abstract Service.

\end{acknowledgements}

\onecolumn
\newpage
\begin{table}
\centering
\caption[]{\label{tab:tab 3} Data sample. Kinematics and evolutionary status.}
\paragraph{}
\begin{tabular}{c c c c c c c c c c c c} 
\multicolumn{12}{c}{}\\
\hline
HD&DM&G&HIP&U&V&W&$\sigma_{\mathrm{U}}$&$\sigma_{\mathrm{V}}$&
$\sigma_{\mathrm{W}}$&kin.&ev.\,s.\\
\hline
HD 400&BD+35 8&{}&HIP 699&-27.&-9.&-9.&4.&8.&4.&disk&to\\
HD 693&BD-16 17&{}&HIP 910&-19.&-13.&-18.&0.&3.&10.&disk&to\\
HD 1581&CPD-65 13&{}&HIP 1599&73.&-4.&-43.&4.&5.&8.&disk&to\\
HD 2454&BD+09 47&{}&HIP 2235&-13.&-31.&-14.&3.&6.&8.&disk&to\\
HD 3567&BD-09 122&G 270-23&HIP 3026&-137.&-236.&-42.&19.&32.&16.&halo&to\\
HD 3823&CD-60 118&{}&HIP 3170&113.&-18.&-33.&3.&4.&8.&disk&to\\
HD 4614&BD+57 150&{}&HIP 3821&30.&-10.&-17.&5.&8.&1.&disk&to\\
HD 4813&BD-11 153&{}&HIP 3909&-21.&-3.&-12.&2.&2.&10.&disk&to\\
HD 5015&BD+60 124&{}&HIP 4151&6.&21.&14.&6.&8.&0.&disk&to\\
HD 6920&BD+41 219&{}&HIP 5493&-35.&8.&-9.&6.&7.&4.&disk&to\\
HD 7439&BD-08 216&{}&HIP 5799&34.&22.&-8.&3.&2.&9.&disk&to\\
HD 7476&BD-01 162&{}&HIP 5833&27.&42.&-4.&3.&3.&9.&disk&to\\
HD 9826&BD+40 332&{}&HIP 7513&-28.&-22.&-15.&6.&7.&3.&disk&to\\
{}&BD+72 94&G 245-32&HIP 8314&-307.&-114.&28.&30.&20.&16.&halo&to\\
HD 11112&CD-42 638&{}&HIP 8398&89.&-43.&-21.&3.&3.&9.&disk&to\\
HD 13555&BD+20 348&{}&HIP 10306&20.&-12.&4.&7.&4.&6.&disk&to\\
HD 14802&CD-24 1038&{}&HIP 11072&19.&-17.&-10.&3.&2.&9.&disk&to\\
HD 15335&BD+29 423&{}&HIP 11548&25.&32.&-14.&7.&5.&5.&disk&to\\
HD 15798&BD-15 449&{}&HIP 11783&-31.&-4.&18.&4.&1.&9.&disk&to\\
HD 16031&BD-13 482&{}&HIP 11952&-11.&-101.&-71.&10.&20.&9.&?&to\\
HD 16895&BD+48 746&{}&HIP 12777&31.&1.&-1.&8.&6.&2.&disk&to\\
HD 17051&CD-51 641&{}&HIP 12653&31.&-17.&-7.&0.&5.&9.&disk&to\\
HD 18768&BD+46 678&{}&HIP 14181&{}&{}&{}&{}&{}&{}&{}&to\\
HD 20407&CD-46 968&{}&HIP 15131&6.&16.&-14.&1.&5.&8.&disk&to\\
HD 20807&CPD-62 265&{}&HIP 15371&70.&-47.&17.&1.&7.&7.&disk&to\\
HD 22484&BD-00 572&{}&HIP 16852&-1.&-15.&-42.&8.&1.&7.&disk&to\\
HD 22879&BD-03 592&G 80-15&HIP 17147&105.&-86.&-40.&7.&2.&7.&disk&to\\
HD 26491&CD-64 143&{}&HIP 19233&39.&-27.&-16.&1.&7.&7.&disk&to\\
HD 284248&BD+21 607&G 8-16&HIP 19797&353.&-147.&-76.&10.&19.&5.&halo&to\\
HD 28620&BD+36 907&{}&HIP 21167&4.&-4.&3.&10.&3.&1.&disk&to\\
HD 30495&BD-17 954&{}&HIP 22263&21.&-6.&0.&7.&5.&6.&disk&to\\
HD 30649&BD+45 992&G 81-38&HIP 22596&57.&-81.&-9.&9.&5.&0.&disk&to\\
HD 32778&CD-56 1071&{}&HIP 23437&76.&13.&-11.&1.&8.&6.&disk&to\\
HD 33256&BD-04 1056&{}&HIP 23941&9.&-6.&3.&8.&4.&4.&disk&to\\
HD 34328&CD-59 1024&{}&HIP 24316&206.&-350.&100.&14.&14.&17.&halo&to\\
HD 34721&BD-18 1051&{}&HIP 24786&36.&-44.&21.&7.&6.&5.&disk&to\\
HD 37655&CD-43 1954&{}&HIP 26488&86.&-22.&22.&4.&8.&5.&disk&to\\
HD 39587&BD+20 1162&{}&HIP 27913&-14.&2.&-7.&10.&2.&0.&disk&to\\
HD 41330&BD+35 1334&{}&HIP 28908&-6.&-25.&-33.&10.&1.&1.&disk&to\\
HD 41640&BD+16 1001&{}&HIP 29001&{}&{}&{}&{}&{}&{}&{}&to\\
HD 43042&BD+19 1270&{}&HIP 29650&33.&-19.&-16.&10.&2.&0.&disk&to\\
HD 43947&BD+16 1091&{}&HIP 30067&39.&-11.&-2.&10.&3.&0.&disk&to\\
HD 48938&CD-27 3248&{}&HIP 32322&24.&26.&17.&5.&8.&2.&disk&to\\
HD 51530&BD+26 1405&{}&HIP 33595&18.&31.&-22.&10.&2.&3.&disk&to\\
HD 53705&CD-43 2906&{}&HIP 34065&52.&-73.&-20.&3.&9.&3.&disk&to\\
HD 55575&BD+47 1419&{}&HIP 35136&80.&-2.&33.&9.&1.&4.&disk&to\\
HD 58551&BD+21 1596&{}&HIP 36152&60.&-4.&-27.&9.&3.&3.&disk&to\\
{}&{}&G 90-3&HIP 36430&{}&{}&{}&{}&{}&{}&{}&sg\\
HD 59984&BD-08 1964&{}&HIP 36640&29.&-51.&-18.&7.&7.&1.&disk&to\\
HD 61421&BD+05 1739&{}&HIP 37279&-5.&-9.&-19.&8.&5.&2.&disk&to\\
HD 62407&BD+13 1750&{}&HIP 37723&23.&-27.&-17.&8.&6.&7.&disk&sg\\
\end{tabular}
\end{table}
\newpage
\begin{table}
\centering
\paragraph{}
\begin{tabular}{c c c c c c c c c c c c}
\multicolumn{12}{c}{}\\
HD 62301&BD+39 1998&{}&HIP 37789&7.&-109.&-22.&9.&4.&5.&?&to\\
HD 63077&CD-33 4113&{}&HIP 37853&145.&-59.&41.&4.&9.&1.&?&to\\
HD 65907&CD-59 1773&{}&HIP 38908&-12.&-23.&34.&1.&10.&3.&disk&to\\
HD 67458&CD-29 5555&{}&HIP 39710&-61.&-6.&10.&4.&9.&0.&disk&to\\
HD 69897&BD+27 1589&{}&HIP 40843&24.&-39.&7.&8.&2.&5.&disk&to\\
HD 73524&CD-39 4574&{}&HIP 42291&27.&6.&-29.&2.&10.&0.&disk&to\\
HD 74000&BD-15 2546&{}&HIP 42592&-249.&-359.&58.&63.&35.&3.&halo&to\\
HD 74011&BD+34 1885&G 115-19&HIP 42734&28.&-70.&24.&8.&3.&6.&disk&to\\
{}&BD+25 1981&G 40-34&HIP 42887&43.&-247.&-94.&8.&51.&28.&halo&to\\
HD 76932&BD-15 2656&{}&HIP 44075&47.&-92.&71.&4.&8.&3.&disk&to\\
HD 79028&BD+62 1058&{}&HIP 45333&-8.&-7.&-9.&7.&3.&7.&disk&to\\
HD 82328&BD+52 1401&{}&HIP 46853&57.&-35.&-24.&7.&2.&7.&disk&to\\
HD 86560&BD+53 1378&{}&HIP 49070&{}&{}&{}&{}&{}&{}&{}&to\\
HD 91347&BD+49 1966&G 196-33&HIP 51700&-50.&28.&-3.&5.&2.&8.&disk&to\\
HD 91752&BD+37 2100&{}&HIP 51914&-20.&-4.&-16.&5.&1.&9.&disk&to\\
HD 91889&BD-11 2918&{}&HIP 51933&-68.&-35.&-36.&2.&8.&6.&disk&to\\
HD 94028&BD+21 2247&G 58-25&HIP 53070&33.&-140.&8.&4.&8.&9.&halo&to\\
HD 95241&BD+43 2068&{}&HIP 53791&11.&-33.&-10.&4.&1.&9.&disk&to\\
HD 97916&BD+02 2406&{}&HIP 55022&-108.&11.&89.&18.&8.&11.&?&to\\
HD 98991&BD-17 3367&{}&HIP 55598&52.&-35.&-18.&2.&8.&6.&disk&to\\
HD 99747&BD+62 1183&{}&HIP 56035&19.&21.&-31.&5.&4.&8.&disk&to\\
HD 102634&BD+00 2843&{}&HIP 57629&29.&-16.&-4.&1.&5.&9.&disk&to\\
HD 103799&BD+41 2253&{}&HIP 58287&32.&-25.&21.&3.&1.&10.&disk&to\\
HD 106516&BD-09 3468&{}&HIP 59750&-54.&-74.&-59.&2.&6.&8.&disk&to\\
HD 107113&BD+87 107&{}&HIP 59879&-37.&14.&4.&5.&7.&5.&disk&to\\
HD 108134&BD+61 1294&{}&HIP 60588&-41.&-6.&-35.&4.&4.&8.&disk&to\\
HD 108177&BD+02 2538&G 13-35&HIP 60632&-111.&-228.&51.&10.&19.&14.&halo&to\\
HD 109358&BD+42 2321&{}&HIP 61317&31.&-4.&2.&2.&2.&10.&disk&to\\
HD 110897&BD+40 2570&{}&HIP 62207&41.&7.&76.&1.&2.&10.&disk&to\\
HD 114762&BD+18 2700&G 63-9&HIP 64426&82.&-70.&59.&6.&4.&10.&disk&to\\
HD 120162&BD+69 717&{}&HIP 67109&{}&{}&{}&{}&{}&{}&{}&to\\
HD 121560&BD+14 2680&{}&HIP 68030&30.&-20.&-3.&4.&0.&9.&disk&to\\
HD 123710&BD+75 526&{}&HIP 68796&36.&-8.&1.&3.&7.&7.&disk&to\\
HD 126512&BD+21 2649&G 166-25&HIP 70520&-85.&-84.&-79.&5.&3.&9.&disk&to\\
HD 128167&BD+30 2536&{}&HIP 71284&-2.&16.&-5.&3.&3.&9.&disk&to\\
HD 131117&CD-30 11780&{}&HIP 72772&59.&-36.&11.&8.&4.&4.&disk&to\\
HD 134169&BD+04 2969&{}&HIP 74079&-3.&-3.&-1.&7.&1.&8.&disk&sg-to\\
{}&{}&G 152-35&HIP 76059&{}&{}&{}&{}&{}&{}&{}&{}\\
HD 141004&BD+07 3023&{}&HIP 77257&49.&-24.&-39.&7.&2.&7.&disk&to\\
HD 142373&BD+42 2648&{}&HIP 77760&42.&11.&-67.&2.&6.&8.&disk&to\\
HD 142860&BD+16 2849&{}&HIP 78072&-56.&-33.&-25.&6.&3.&7.&disk&to\\
HD 143761&BD+33 2663&{}&HIP 78459&-55.&-35.&21.&4.&5.&7.&disk&to\\
{}&BD+42 2667&G 180-24&HIP 78640&-109.&-268.&-28.&21.&25.&14.&halo&to\\
HD 146588&BD+19 3076&{}&HIP 79720&{}&{}&{}&{}&{}&{}&{}&to\\
HD 148816&BD+04 3195&G 17-21&HIP 80837&-83.&-264.&-81.&9.&10.&6.&halo&to\\
HD 150453&BD-19 4406&{}&HIP 81754&-5.&10.&2.&10.&1.&3.&disk&to\\
HD 155358&BD+33 2840&{}&HIP 83949&{}&{}&{}&{}&{}&{}&{}&to\\
HD 157089&BD+01 3421&{}&HIP 84905&167.&-42.&-9.&9.&4.&4.&halo&to\\
HD 159332&BD+19 3354&{}&HIP 85912&28.&-48.&-26.&7.&6.&4.&disk&to\\
HD 160291&BD+48 2541&{}&HIP 86173&-34.&8.&-25.&3.&8.&6.&disk&to\\
{}&BD+18 3423&G 170-56&HIP 86321&83.&-265.&-50.&15.&20.&9.&halo&to\\
HD 160693&BD+37 2926&G 182-19&HIP 86431&-209.&-112.&85.&9.&10.&6.&halo&to\\
HD 160617&CD-40 11755&{}&HIP 86694&-59.&-216.&-94.&12.&29.&12.&halo&sg\\
{}&BD-08 4501&G 20-15&HIP 87062&-141.&-49.&-163.&15.&17.&38.&halo&to\\
\end{tabular}
\end{table}
\newpage
\begin{table}
\centering
\paragraph{}
\begin{tabular}{c c c c c c c c c c c c} 
\multicolumn{12}{c}{}\\
HD 165908&BD+30 3128&{}&HIP 88745&6.&1.&10.&5.&8.&4.&disk&to\\
HD 166913&CD-59 6824&{}&HIP 89554&45.&-48.&69.&9.&6.&5.&disk&to\\
HD 167588&BD+29 3213&{}&HIP 89408&-41.&-17.&-17.&5.&8.&3.&disk&to\\
HD 168151&BD+64 1252&{}&HIP 89348&6.&-13.&-51.&1.&9.&5.&disk&to\\
HD 170153&BD+72 839&{}&HIP 89937&-3.&40.&-3.&2.&9.&5.&disk&to\\
HD 174912&BD+38 3327&G 207-5&HIP 92532&22.&8.&-42.&4.&9.&3.&disk&to\\
HD 181743&CD-45 13178&{}&HIP 95333&47.&-334.&-60.&13.&52.&9.&halo&to\\
HD 186379&BD+24 3849&{}&HIP 97023&-32.&-27.&-46.&5.&9.&2.&disk&to\\
HD 187691&BD+10 4073&{}&HIP 97675&3.&-3.&-25.&6.&8.&2.&disk&to\\
{}&{}&G 24-3&HIP 98989&{}&{}&{}&{}&{}&{}&{}&to\\
HD 345957&BD+23 3912&{}&HIP 99423&{}&{}&{}&{}&{}&{}&{}&to\\
HD 194598&BD+09 4529&G 24-15&HIP 100792&74.&-276.&-30.&7.&10.&7.&halo&to\\
HD 195633&BD+06 4557&{}&HIP 101346&35.&8.&-24.&7.&8.&5.&disk&sg-to\\
HD 199288&CD-44 14214&{}&HIP 103458&-22.&-101.&46.&8.&2.&7.&disk&to\\
HD 199960&BD-05 5433&{}&HIP 103682&7.&-24.&-3.&6.&6.&5.&disk&to\\
HD 200580&BD+02 4295&G 25-15&HIP 103987&-96.&-75.&9.&9.&9.&5.&disk&sg\\
HD 201891&BD+17 4519&{}&HIP 104659&-92.&-115.&-59.&5.&9.&4.&?&to\\
HD 202628&CD-43 14464&{}&HIP 105184&12.&2.&-27.&7.&0.&7.&disk&to\\
HD 203454&BD+39 4529&{}&HIP 105406&-20.&-2.&-17.&1.&10.&1.&disk&to\\
HD 205650&CD-28 17381&{}&HIP 106749&118.&-83.&11.&7.&5.&9.&disk&to\\
HD 207129&CD-47 13928&{}&HIP 107649&13.&-22.&1.&6.&1.&8.&disk&to\\
HD 207978&BD+28 4215&{}&HIP 107975&-13.&16.&-8.&1.&9.&3.&disk&to\\
HD 208906&BD+29 4550&{}&HIP 108490&-73.&-2.&-12.&2.&9.&3.&disk&to\\
{}&BD+17 4708&G 126-62&HIP 109558&302.&-283.&11.&42.&10.&24.&halo&to\\
HD 210918&CD-41 14804&{}&HIP 109821&47.&-92.&-9.&6.&1.&8.&disk&to\\
HD 211415&CD-54 9222&{}&HIP 110109&30.&-41.&7.&6.&2.&8.&disk&to\\
{}&BD+07 4841&G 18-39&HIP 110140&267.&-317.&-107.&107.&74.&128.&halo&to\\
HD 212698&BD-17 6521&{}&HIP 110778&19.&-6.&-5.&4.&4.&8.&disk&to\\
{}&{}&G 18-54&HIP 111195&-11.&-267.&51.&15.&32.&24.&halo&to\\
HD 214953&CD-47 14307&{}&HIP 112117&-14.&-38.&-6.&5.&1.&9.&disk&to\\
HD 216385&BD+09 5122&{}&HIP 112935&58.&-7.&-33.&2.&7.&7.&disk&to\\
HD 218470&BD+48 3944&{}&HIP 114210&30.&-9.&10.&3.&9.&2.&disk&to\\
HD 218502&BD-15 6355&{}&HIP 114271&{}&{}&{}&{}&{}&{}&{}&to\\
HD 219476&BD+30 4912&{}&HIP 114838&{}&{}&{}&{}&{}&{}&{}&to\\
HD 219623&BD+52 3410&{}&HIP 114924&-7.&-27.&-22.&3.&9.&1.&disk&to\\
HD 219617&BD-14 6437&G 273-1&HIP 114962&-383.&-329.&-58.&76.&67.&13.&halo&to\\
{}&BD+02 4651&G 29-23&HIP 115167&299.&-453.&-60.&192.&206.&172.&halo&sg\\
{}&BD+59 2723&G 217-8&HIP 115704&180.&-186.&-55.&36.&18.&9.&halo&to\\
HD 221377&BD+51 3630&{}&HIP 116082&{}&{}&{}&{}&{}&{}&{}&sg\\
HD 222368&BD+04 5035&{}&HIP 116771&8.&-27.&-26.&0.&6.&8.&disk&to\\
\hline
\end{tabular}
\end{table}
\begin{table}
\centering
\caption[]{\label{tab:tab 4} Data sample. Effective temperatures, gravities, 
metallicities and lithium abundances as taken from the literature. The 
metallicities given in the brackets are those derived from the photometry, 
all the others refer to spectroscopic determinations. References: 
1 = Deliyannis et al.\,, 1990 (their SS); 2 = Deliyannis et al.\,, 1990 (their 
HD); 3 = Deliyannis et al.\,, 1990 (their RMB); 4 = Lambert et al.\,, 1991; 
5 = Pilachowski et al.\,, 1993; 6 = Pasquini et al.\,, 1994; 7 = Spite et 
al.\,, 1996 (in this paper the temperature has been determined from either 
the excitation balance of the FeI lines or the dereddened $(b - y)_0$ color 
or the profile of the H$_\alpha$ wings. The gravity has been estimated from 
the position in the $c_1$\,--\,$(b - y)_0$ diagram or by comparing the iron 
abundance deduced from FeI and FeII lines. When multiple determinations are 
given, we list all of them).}
\paragraph{}
\begin{tabular}{c c c c c c c c}
\multicolumn{8}{c}{}\\\hline
HIP&T$_{\mathrm{eff}}$&log g&[Fe/H]&W(Li)&A(Li)&upper limits&ref.\\
\hline
HIP 699&6190&4.13&-0.35&22&+2.18&{}&4\\
HIP 910&6200&4.07&-0.38&34&+2.39&{}&4\\
HIP 1599&6009&4.52&-0.26 (-0.15)&40&+2.36&{}&6\\
HIP 2235&6490&4.08&-0.37&4&+1.62&u&4\\
HIP 3026&5950&4.00&-1.20&45&+2.35&{}&3\\
HIP 3170&6037&4.34&-0.35 (-0.35)&45&+2.44&{}&6\\
HIP 3821&5950&4.47&-0.31&21&+1.98&{}&4\\
HIP 3909&6250&4.32&-0.15&64&+2.77&{}&4\\
HIP 4151&6200&3.98&+0.00&2&+1.10&u&4\\
HIP 5493&5800&3.88&-0.21&4&+1.13&u&4\\
HIP 5799&6470&4.10&-0.32&4&+1.58&{}&4\\
HIP 5833&6520&4.01&-0.24&29&+2.54&{}&4\\
HIP 7513&6210&4.17&+0.09&30&+2.36&{}&4\\
HIP 8314&6160&4.50&-1.80&27&+2.22&{}&3\\
{}&6160&3.00&{}&27.0&+2.22&{}&5\\
HIP 8398&5800&4.03&-0.07 (-0.20)&57&+2.37&{}&6\\
HIP 10306&6360&4.07&-0.32&64&+2.84&{}&4\\
HIP 11072&5905&4.19&-0.19 (-0.07)&51&+2.40&{}&6\\
HIP 11548&5860&4.06&-0.22&52&+2.34&{}&4\\
HIP 11783&6440&3.94&-0.25&81&+3.03&{}&4\\
HIP 11952&5929&4.00&-2.20&28.0&+2.03&{}&1\\
{}&5970&3.90&-1.89 (-1.89)&28.0&+2.07&{}&7\\
HIP 12777&6310&4.30&-0.02&67&+2.84&{}&4\\
HIP 12653&6074&4.22&-0.04 (+0.01)&38&+2.39&{}&6\\
HIP 14181&5720&4.04&-0.62&61&+2.31&{}&4\\
HIP 15131&5879&4.32&-0.55 (-0.37)&15&+1.79&{}&6\\
HIP 15371&5856&4.40&-0.38 (-0.50)&3&+0.93&u&6\\
HIP 16852&5980&4.15&-0.15&41&+2.33&{}&4\\
HIP 17147&5740&4.10&-0.60&25&+1.80&{}&2\\
HIP 19233&5744&4.19&-0.28 (-0.30)&3&+0.93&u&6\\
HIP 19797&5929&4.00&-1.60&25.0&+1.98&{}&1\\
HIP 21167&6140&4.06&(-0.61)&34&+2.34&{}&4\\
HIP 22263&5829&4.30&-0.13 (+0.01)&56&+2.39&{}&6\\
HIP 22596&5700&4.10&-0.30&33&+1.90&{}&2\\
{}&5740&4.22&-0.51&2&+0.76&u&4\\
HIP 23437&5760&4.34&-0.61 (-0.30)&8&+1.40&{}&6\\
HIP 23941&6440&4.05&-0.30&3&+1.43&u&4\\
HIP 24316&5730&4.60&-1.60  (-1.60)&32.0&+1.98&{}&7\\
HIP 24786&6001&4.09&-0.25 (-0.18)&39&+2.35&{}&6\\
HIP 26488&5874&4.04&-0.31 (-0.31)&46&+2.32&{}&6\\
HIP 27913&5950&4.46&-0.03&95&+2.77&{}&4\\
HIP 28908&5920&4.14&-0.24&19&+1.89&{}&4\\
HIP 29001&6080&4.33&(-0.71)&49&+2.48&{}&4\\
HIP 29650&6590&4.27&+0.04&6&+1.86&{}&4\\
HIP 30067&5950&4.28&-0.30&38&+2.25&{}&4\\
HIP 32322&6018&4.30&-0.56 (-0.59)&33&+2.27&{}&6\\
HIP 33595&6020&3.94&-0.56&3&+1.16&u&4\\
HIP 34065&5812&4.33&-0.36 (-0.26)&3&+0.98&u&6\\
\end{tabular}
\end{table}
\newpage
\begin{table}
\centering
\paragraph{}
\begin{tabular}{c c c c c c c c}
\multicolumn{8}{c}{}\\
HIP 35136&5960&4.48&-0.28&10&+1.62&{}&4\\
HIP 36152&6180&4.17&-0.55&36&+2.40&{}&4\\
HIP 36430&5900&3.00&-1.93&43.5&+2.25&{}&5\\
HIP 36640&5980&4.18&-0.74&43&+2.33&{}&4\\
HIP 37279&6700&4.03&-0.02&2&+1.48&u&4\\
HIP 37723&5820&4.26&(-0.71)&18&+1.78&{}&4\\
HIP 37789&5900&4.19&-0.69&27&+2.03&{}&4\\
HIP 37853&5778&4.27&-0.90 (-0.78)&8&+1.41&{}&6\\
HIP 38908&6072&4.50&-0.36 (-0.20)&3&+1.20&u&6\\
HIP 39710&5962&4.47&-0.24 (-0.06)&26&+2.12&{}&6\\
HIP 40843&6360&4.35&-0.26&52&+2.73&{}&4\\
HIP 42291&5972&4.32&-0.01 (+0.06)&46&+2.41&{}&6\\
HIP 42592&6223&4.50&-1.80&24.5&+2.16&{}&1\\
HIP 42734&5740&4.15&-0.65&16&+1.69&{}&4\\
HIP 42887&6780&4.10&-1.30&12&+2.50&u&2\\
HIP 44075&5861&3.50&-1.10&23.0&+1.96&{}&1\\
{}&5970&4.37&-0.82&24&+2.05&{}&4\\
{}&5900&3.00&-0.90&26.7&+2.02&{}&5\\
HIP 45333&5880&4.18&-0.08&11&+1.63&{}&4\\
HIP 46853&6380&4.09&-0.20&102&+3.15&{}&4\\
HIP 49070&5910&4.13&(-0.59)&28&+2.07&{}&4\\
HIP 51700&5870&4.24&-0.48&21&+1.91&{}&4\\
HIP 51914&6490&3.92&-0.23&2&+1.33&u&4\\
HIP 51933&6140&4.22&-0.24&38&+2.41&{}&4\\
HIP 53070&5794&4.00&-1.70&35.0&+2.09&{}&1\\
{}&5860&4.10&-1.70&33&+2.10&{}&2\\
{}&5800&3.00&-1.51&36.6&+2.09&{}&5\\
HIP 53791&5890&4.05&-0.30&1&+0.35&u&4\\
HIP 55022&6124&4.00&-1.10&3.0&+1.20&u&1\\
{}&6000&3.00&-1.38&5.0&+1.30&u&5\\
HIP 55598&6640&3.98&-0.11&3&+1.63&u&4\\
HIP 56035&6610&3.99&-0.54&5&+1.80&u&4\\
HIP 57629&6390&4.18&+0.24&33&+2.53&{}&4\\
HIP 58287&6220&4.04&(-0.43)&19&+2.13&{}&4\\
HIP 59750&6067&4.30&-0.40&11.0&+1.80&u&1\\
{}&6110&4.10&-0.90&6&+1.50&u&2\\
{}&6250&4.38&-0.70&3&+1.32&u&4\\
HIP 59879&6390&4.07&-0.54&4&+1.55&u&4\\
HIP 60588&5830&4.21&(-0.58)&20&+1.85&{}&4\\
HIP 60632&5847&4.50&-1.90&35.0&+2.09&{}&1\\
{}&5900&4.10&-1.90&24&+1.90&u&2\\
{}&5861&{}&-1.90&29&+2.00&{}&3\\
HIP 61317&5880&4.52&-0.19&9&+1.53&{}&4\\
HIP 62207&5794&3.90&-0.30&33.0&+2.10&{}&1\\
{}&5800&4.15&-0.59&20&+1.82&{}&4\\
HIP 64426&5750&4.10&-0.80&17&+1.60&{}&2\\
{}&5740&4.00&-0.80&25&+1.88&{}&3\\
{}&5870&4.24&-0.74&22&+1.92&{}&4\\
HIP 67109&5900&4.38&(-0.80)&31&+2.11&{}&4\\
HIP 68030&6190&4.36&(-0.37)&38&+2.44&{}&4\\
HIP 68796&5740&4.00&-0.60&23&+1.85&{}&3\\
HIP 70520&5750&4.20&-0.63&16&+1.69&{}&4\\
HIP 71284&6770&4.27&-0.41&3&+1.67&u&4\\
HIP 72772&6000&4.09&+0.13&66&+2.60&{}&4\\
\end{tabular}
\end{table}
\newpage
\begin{table}
\centering
\paragraph{}
\begin{tabular}{c c c c c c c c}
\multicolumn{8}{c}{}\\
HIP 74079&5794&3.80&-1.60&44.0&+2.23&{}&1\\
{}&5800&4.10&-1.60&46&+2.20&{}&2\\
{}&5800&3.00&{}&44.0&+2.18&{}&5\\
HIP 76059&5600, 5700&4.50, 3.80&-1.75 (-1.61)&53.7&+2.13, +2.21&{}&7\\
HIP 77257&5940&4.21&-0.04&20&+1.95&{}&4\\
HIP 77760&5794&3.90&-0.30&69.0&+2.58&{}&1\\
{}&5830&4.10&-0.40&56&+2.40&{}&2\\
{}&5840&4.34&-0.52&49&+2.30&{}&4\\
HIP 78072&6330&4.25&-0.16&17&+2.15&{}&4\\
HIP 78459&5780&4.24&-0.26&6&+1.28&{}&4\\
HIP 78640&5960&4.00&-1.70&28&+2.12&{}&3\\
HIP 79720&5980&4.33&(-0.73)&38&+2.40&{}&4\\
HIP 80837&5534&4.00&-0.50&17.0&+1.53&{}&1\\
{}&5810&4.10&-0.70&18&+1.70&{}&2\\
HIP 81754&6440&3.86&-0.31&29&+2.49&{}&4\\
HIP 83949&5870&4.19&-0.67&20&+1.87&{}&4\\
HIP 84905&5861&3.70&-0.50&25.0&+2.11&{}&1\\
{}&5735&4.10&-0.60&20&+1.70&{}&2\\
HIP 85912&6240&3.91&-0.23&4&+1.45&u&4\\
HIP 86173&6070&4.19&(-0.61)&25&+2.14&{}&4\\
HIP 86321&6067&4.00&-1.00&31.0&+2.19&{}&1\\
{}&6140&4.00&-0.80&34&+2.35&{}&3\\
HIP 86431&5780&4.10&-0.70&7&+1.20&u&2\\
{}&5710&4.00&-0.70&4&+1.00&u&3\\
HIP 86694&5861&3.50&-1.60&42.0&+2.20&{}&1\\
{}&5900&3.00&{}&42.0&+2.23&{}&5\\
HIP 87062&5600&3.00&-1.78&28.3&+1.80&{}&5\\
{}&5830, 5900&4.00, 4.00&-1.40 (-1.58)&35.1&+2.09, +2.14&{}&7\\
HIP 88745&5998&4.20&-0.40&41.0&+2.38&{}&1\\
{}&6020&4.48&-0.56&34&+2.25&{}&4\\
HIP 89554&5861&3.30&-1.80&40.0&+2.17&{}&1\\
HIP 89408&5900&4.23&(-0.38)&41&+2.26&{}&4\\
HIP 89348&6590&4.09&-0.32&2&+1.40&u&4\\
HIP 89937&6152&4.30&-0.30&42.0&+2.52&{}&1\\
{}&5920&4.10&-0.30&29&+2.00&{}&2\\
HIP 92532&5860&4.33&-0.54&29&+2.05&{}&4\\
HIP 95333&5790, 5900&4.50, 3.50&-1.70 (-1.89)&40.6&+2.13, +2.21&{}&7\\
HIP 97023&5880&4.11&(-0.55)&40&+2.22&{}&4\\
HIP 97675&6150&4.14&+0.09&63&+2.69&{}&4\\
HIP 98989&5800&{}&-1.63&28.5&+1.97&{}&5\\
HIP 99423&5596&4.00&-1.70&67.0&+2.23&{}&1\\
{}&5720&4.00&-1.30&73&+2.40&{}&3\\
{}&5600&3.00&-1.59&80.8&+2.38&{}&5\\
HIP 100792&5808&4.00&-1.60&29.0&+2.00&{}&1\\
HIP 101346&5840&3.80&-0.90&40&+2.15&{}&3\\
HIP 103458&5727&4.35&-0.82 (-0.64)&3&+0.86&u&6\\
HIP 103682&5810&4.20&+0.11&53&+2.32&{}&4\\
HIP 103987&5730&3.50&-0.70&32&+2.00&{}&3\\
HIP 104659&5794&4.50&-1.40&23.0&+1.89&{}&1\\
{}&5810&4.10&-1.40&27&+1.90&{}&2\\
{}&5850&4.50&-1.00&27&+2.08&{}&3\\
{}&5870&4.46&-1.06&21&+1.90&{}&4\\
HIP 105184&5771&4.52&-0.14&41&+2.17&{}&6\\
HIP 105406&5750&4.10&-0.30&64&+2.40&{}&2\\
\end{tabular}
\end{table}
\newpage
\begin{table}
\centering
\paragraph{}
\begin{tabular}{c c c c c c c c}
\multicolumn{8}{c}{}\\
HIP 106749&5728&4.00&-1.30&18.0&+1.67&{}&1\\
HIP 107649&5948&4.13&-0.15 (-0.05)&38&+2.29&{}&6\\
HIP 107975&5780&4.10&-0.50&5&+1.10&u&2\\
HIP 108490&5900&4.10&-0.50&43&+2.20&{}&2\\
{}&5960&4.00&-0.90&50&+2.40&{}&3\\
{}&6010&4.41&-0.72&38&+2.30&{}&4\\
HIP 109558&5810&4.10&-2.00&25&+1.90&{}&2\\
{}&5890&4.00&-1.70&25&+1.98&{}&3\\
HIP 109821&5802&4.43&-0.18 (+0.10)&3&+0.94&u&6\\
HIP 110109&5870&4.35&-0.36 (-0.32)&10&+1.60&{}&6\\
HIP 110140&5920&4.00&-1.20&37&+2.24&{}&3\\
HIP 110778&5915&4.50&-0.13 (-0.14)&114&+2.89&{}&6\\
HIP 111195&5800&3.00&-1.33&34.9&+2.07&{}&5\\
HIP 112117&6069&4.49&-0.09 (+0.07)&65&+2.67&{}&6\\
HIP 112935&6067&3.90&-0.62&47.0&+2.53&{}&1\\
{}&6290&3.97&-0.25&2&+1.28&u&4\\
HIP 114210&6600&4.21&-0.13&4&+1.72&u&4\\
HIP 114271&6000&3.00&-1.96&30.0&+2.16&{}&5\\
HIP 114838&5930&4.01&(-0.68)&2&+0.91&u&4\\
HIP 114924&6130&4.21&+0.00&64&+2.68&{}&4\\
HIP 114962&5662&3.90&-1.40&42.0&+2.04&{}&1\\
{}&5820&4.10&-1.40&43&+2.20&{}&2\\
{}&5870&4.50&-1.50&40&+2.20&{}&3\\
HIP 115167&5794&4.00&-2.30&27.0&+1.92&{}&1\\
HIP 115704&5830&4.00&-1.60&25&+1.95&{}&3\\
HIP 116082&6000&3.50&-1.10&6&+1.45&u&3\\
{}&6320&3.89&(-0.72)&3&+1.41&u&4\\
HIP 116771&5998&3.90&-0.50&22.0&+2.05&{}&1\\
{}&6260&4.16&-0.17&18&+2.14&{}&4\\
\hline
\end{tabular}
\end{table}
\twocolumn

\end{document}